\documentclass[prd,twocolumn]{revtex4}

\pdfoutput=1

\usepackage{graphicx}
\usepackage{dcolumn}
\usepackage{bm}
\usepackage{amssymb,amsmath,bm}  
\usepackage{color}
\usepackage{hyperref}
\usepackage{multirow}
\usepackage[utf8]{inputenc}
\usepackage{balance}
\usepackage{enumitem}
\usepackage{lipsum}
\newcommand{\uKam}{\mu\text{K-arcmin}}
\newcommand{\nv}{\hat{\bf n}}

\begin{document}
\title{Simulated forecasts for primordial $B$-mode searches in ground-based experiments}
\author{David Alonso$^1$, Joanna Dunkley$^1$, Sigurd N\ae ss$^1$, Ben Thorne$^1$}
\affiliation{$^{1}$University of Oxford, Denys Wilkinson Building,
             Keble Road, Oxford, OX1 3RH,  UK}

\begin{abstract}
  Detecting the imprint of inflationary gravitational waves on the $B$-mode polarization
  of the Cosmic Microwave Background (CMB) is one of the main science cases for current
  and next-generation CMB experiments. In this work we explore some of the challenges
  that ground-based facilities will have to face in order to carry out this measurement
  in the presence of Galactic foregrounds and correlated atmospheric noise. We present
  forecasts for Stage-3 (S3) and planned Stage-4 (S4) experiments based on the analysis
  of simulated sky maps using a map-based Bayesian foreground cleaning method. Our
  results thus consistently propagate the uncertainties on foreground parameters such
  as spatially-varying spectral indices, as well as the bias on the measured
  tensor-to-scalar ratio $r$ caused by an incorrect modelling of the foregrounds.
  We find that S3 and S4-like experiments should be able to put constraints on $r$
  of the order $\sigma(r)=(0.5-1.0)\times10^{-2}$ and $\sigma(r)=(0.5-1.0)\times10^{-3}$
  respectively, assuming instrumental systematic effects are under control. We further study
  deviations from the fiducial foreground model, finding that, while the effects of a
  second polarized dust component would be minimal on both S3 and S4, a 2\% polarized
  anomalous dust emission (AME) component would be clearly detectable by Stage-4
  experiments.
\end{abstract}

  \date{\today}
  \maketitle

\section{Introduction}\label{sec:intro}
  The CMB $B$-mode polarization signal contains a wealth of information on
  both the physics of the primordial Universe, through the unequivocal signal
  of gravitational waves generated during inflation \cite{1997PhRvL..78.2058K,
  1997PhRvL..78.2054S}, as well as on the late-time evolution of the Universe,
  through the distortion of $E$-mode polarization caused by gravitational lensing
  \cite{1998PhRvD..58b3003Z}. A detection of the former would not only strengthen
  the position of the inflationary hypothesis, but also effectively measure the
  energy scale of inflation. Constraining this quantity below the level
  of $r\sim10^{-2}-10^{-3}$ would be of tremendous importance for inflationary
  theories \cite{1997PhRvL..78.1861L}. For these reasons, significant effort has
  been put by the CMB community into building experiments sensitive enough to
  measure this signal, and the first detections of the lensed $B$-mode signal
  have recently started to appear \cite{2014ApJ...794..171T,2015ApJ...807..151K,
  2014JCAP...10..007N}. However, the first attempts to measure the primordial
  signal \cite{2014PhRvL.112x1101B} have been limited by the presence of high
  polarized Galactic foregrounds, in particular polarized thermal dust emission
  \cite{2015PhRvL.114j1301B,2016A&A...586A.133P}.
  
  The challenge of measuring the primordial $B$-mode polarization signal is
  therefore strongly dependent on our ability to disentangle the different sky
  components. Accurate models for the spectral properties of both signal and
  foregrounds must be developed in order
  to optimally separate both components and yield a robust measurement of the
  $B$-mode power spectrum on degree-scales $\ell\lesssim200$. The wide angular
  and frequency coverage afforded by space-borne missions would therefore make
  these experiments ideal for $B$-mode measurements 
  (e.g \cite{2011arXiv1102.2181T,2011JCAP...07..025K,2014JLTP..176..733M}). In
  practice, however, the high cost of space missions, together with the higher
  angular resolution achievable from large ground-based telescopes, has motivated
  the design of several highly competitive sub-orbital facilities. These experiments
  must, nevertheless, cope with a number of limitations, such as the potentially
  large atmospheric systematics on large angular scales or the reduced number of
  atmospheric frequency bands in which CMB observations can be carried out. The
  latter factor can have a significant impact on an experiment's ability to
  separate signal and foregrounds, while the former makes it hard to reliably
  measure some of the most important large-scale features of the primordial
  $B$-mode signal, such as the reionization bump at $\ell\sim10$.
  
  In this work we will study the ability of ground-based experiments to measure
  primordial $B$-modes in the presence of these difficulties. Similar forecasts
  for space and balloon-borne missions have been presented by \cite{2013JCAP...04..047F,
  2016MNRAS.458.2032R}, and a general forecasting framework in the context of
  the Fisher matrix approximation was presented in \cite{2016JCAP...03..052E},
  including a consistent treatment of delensing.
  
  Our methodology to produce these forecasts consists of generating sky simulations
  containing both the cosmological signal as well as realistic Galactic foregrounds
  spanning a range of plausible models. Each simulation is then run through a
  Bayesian component-separation algorithm followed by a power-spectrum estimator,
  with the aim of mimicking as closely as possible the analysis pipeline that real
  observations would be subjected to. This way we can robustly quantify the potential
  bias on $r$ caused by an incorrect modelling of the foregrounds, consistently
  propagate the uncertainties on foreground spectral parameters, and study the 
  needs of these experiments in terms of frequency and area coverage.
  
  The paper is structured as follows: Section \ref{sec:method} describes the
  method, including the models used in the sky simulations, the
  map-based Bayesian component-separation algorithm and the estimator used to
  obtain a measurement of the tensor-to-scalar ratio from the angular power
  spectrum of the foreground-clean map. Section \ref{sec:results} then presents
  the main results of the paper, starting with a simple Fisher forecast in the
  ideal case of flat noise levels and homogeneous foreground spectral parameters.
  We then study the degradation in the constraints on $r$ when accounting for
  spatially-varying spectral indices and in the presence of correlated large-scale
  noise. Making use of more complex foreground simulations, we then quantify the bias
  on $r$ induced by an incorrect modelling of the foregrounds. Finally we
  compare the results of this method with those of a blind foreground cleaning
  algorithm, in order to evaluate the robustness of our findings.
  Section \ref{sec:discuss} summarizes the main results of the papers, and a
  number of technical details regarding the Bayesian component-separation
  algorithm are discussed in Appendices \ref{app:methods} and \ref{app:volume}.
  
\section{Method}\label{sec:method}
  \subsection{Simulations}\label{ssec:method_sims}
    \begin{table*}
    \centering{
      \begin{tabular}{cccccccccc}
      \hline
      Cleaning method & $N_{\rm side}^\beta$ & $B_{\rm NILC}$ & Thermal dust & AME & 
      $\ell_{\rm knee}$ & De-lensing & Area ($10^3\,{\rm deg}^2$) & Experiment & $\#$ sims\\
      \hline
      Bayesian & 4, 8, 16  & N$/$A  & 1 comp. & None       & 0       & w., w.o. & 16, 8, 4, 2 & S3 & 24 \\
      Bayesian & 8, 16, 32 & N$/$A  & 1 comp. & None       & 0       & w., w.o. & 16, 8, 4, 2 & S4 & 24 \\
      Bayesian & 4, 8, 16  & N$/$A  & 1 comp. & None       & 50, 100 & w.       & 4           & S3 & 6  \\
      Bayesian & 8, 16, 32 & N$/$A  & 1 comp. & None       & 50, 100 & w.       & 4           & S4 & 6  \\
      Bayesian & 4, 8, 16  & N$/$A  & 2 comp. & None       & 0       & w., w.o. & 16, 8, 4, 2 & S3 & 24 \\
      Bayesian & 8, 16, 32 & N$/$A  & 2 comp. & None       & 0       & w., w.o. & 16, 8, 4, 2 & S4 & 24 \\
      Bayesian & 4, 8, 16  & N$/$A  & 1 comp. & $2\%$ pol. & 0       & w., w.o. & 16, 8, 4, 2 & S3 & 24 \\
      Bayesian & 8, 16, 32 & N$/$A  & 1 comp. & $2\%$ pol. & 0       & w., w.o. & 16, 8, 4, 2 & S4 & 24 \\
      NILC     & N$/$A     & 1.5, 2 & 1 comp. & None       & 0       & w.       & 16, 8, 4, 2 & S3 & 8  \\
      \hline
    \end{tabular}}
    \caption{Summary of the different simulations run in this work. The column $N_{\rm side}^\beta$
             shows the size of pixels over which the spectral indices are assumed to be constant
             in our Bayesian cleaning approach, while the quantity $B_{\rm NILC}$ determines the
             number of needlet coefficients (and their resolution in $\ell$-space) used in the
             NILC analysis. }
    \label{tab:sims}
    \end{table*}
    In order to study the effect of foregrounds on CMB $B$-mode searches we have
    generated simulations of the observed sky that include the most relevant
    components. For this we use the code {\tt PySM} \cite{pysm}, including the
    following components:
    \begin{enumerate}
      \item {\bf CMB}: the CMB primary anisotropies are straightforward to simulate as
            Gaussian random realizations for a particular power spectrum. We obtained
            this power spectrum from the Boltzmann code CAMB \cite{2000ApJ...538..473L},
            using as input the best-fit cosmological parameters of
            \cite{2015arXiv150201589P} with a tensor-to-scalar ratio $r=0$.
            
            Besides the primary anisotropies, it is also important to include the
            effect of CMB lensing, which generates a $B$-mode signal from the
            associated $E$-$B$ leakage. CMB lensing is a second order effect and
            is therefore non-Gaussian and harder to simulate. For this
            {\tt PySM} uses the algorithm presented in \cite{2013JCAP...09..001N}, which lenses
            the primordial anisotropies given a realization of the lensing potential $\phi$
            using a Taylor expansion of the displaced anisotropies around the
            position of the nearest pixel.
                        
            CMB lensing represents a problem for $B$-mode searches in that it drowns
            the primordial signal by lifting the amplitude of the $BB$ power spectrum
            (and consequently the cosmic-variance contribution to the statistical
            uncertainties). However, given an external estimate of the lensing
            potential it is in principle possible to ``delens'' the $B$-modes
            \cite{2002PhRvL..89a1303K,2002PhRvL..89a1304K,2003PhRvD..67h3002O,
            2012JCAP...06..014S}, thus reducing the final uncertainties on $r$.
            Here we have simulated the effects of delensing simply as a constant
            efficiency factor $f_{\rm dl}$, multiplying the power spectrum of the
            lensing potential:
            \begin{equation}
              C^{\phi\phi,\,delens}_\ell=f_{\rm dl}\,C^{\phi\phi}_\ell
            \end{equation}
            (with the cross-power spectra $C^{T\phi}_\ell$ and $C^{E\phi}_\ell$
            multiplied by $\sqrt{f_{\rm dl}}$).            
            The delensing factor used here depends on the map-level noise of the
            experiment, and was determined using the results of \cite{2016JCAP...03..052E}.
            
            As noted above, the model used in our simulations assumes no primordial
            $B$-modes ($r=0$). The final uncertainties on $r$ would however increase
            with respect to the values found in this paper if the true value of
            $r$ was non-zero, caused by the corresponding non-negligible cosmic
            variance. Thus, the uncertainties quoted in this work correspond to the
            smallest value of $r$ that would be possible to discard at the $1\sigma$
            level.
            
            Since the focus of this work is large-scale primordial $B$-modes, we have not
            included other secondary anisotropies (e.g. Sunyaev-Zel'dovich) or contamination
            from extragalactic sources (e.g. the cosmic infrared backgorund or point
            sources).
            
      \item {\bf Synchrotron}: Galactic synchrotron radiation is caused by cosmic-ray
            electrons interacting with the Galactic magnetic field, and is
            characterized by a smooth power-law-like spectral dependence (see
            \cite{1986rpa..book.....R} for an extended discussion of the physical principles
            of synchrotron emission). Synchrotron is the most important polarized
            foreground at low frequencies. The effects of Faraday rotation in the
            spectral dependence of polarized synchrotron are negligible at the typical
            frequencies of CMB experiments, and we will therefore ignore them here. 
            
            In order to simulate Galactic synchrotron, {\tt PySM} uses a process similar to
            the one used in the design of the Planck Sky Model \cite{2013A&A...553A..96D}.
            The code generates degree-scale templates in intensity for the amplitude and spectral
            index based on the observed variations in sky temperature between the 408 MHz
            Haslam map \cite{1982A&AS...47....1H} and the WMAP maps \cite{2013ApJS..208...20B},
            using the estimate from \cite{2013A&A...553A..96D}.
            
            We add small-scale structure to the large-scale amplitude template through
            a procedure similar to that used in \cite{2007A&A...469..595M,2013A&A...553A..96D}.
            First, we extrapolate the power spectrum of the large-scale template to smaller
            scales as a power-law $C_\ell\propto\ell^{\alpha_s}$, with $\alpha_s=-2.7$.
            We then generate a Gaussian realization of this power spectrum and apply a
            high-pass filter on it to suppress its power on scales already constrained by
            the large-scale template. The amplitude of this small-scale contribution is
            further modulated spatially by multiplying it by a power of the normalized
            local mean intensity of the large-scale template smoothed on scales
            $\theta_{\rm sm}=10^\circ$. The small-scale fluctuations are then added to
            the large-scale amplitude template, and then scaled to different frequencies
            using the spectral index template assuming a perfect power-law behavior.
            Explicitly, the model used for these simulations is:
            \begin{equation}
              T_{\rm sync}(\nu,\nv)=
              \left[T_{\rm LS}(\nu_0,\nv)+
              A\left(\frac{T_{10^\circ}(\nv)}{\bar{T}_{10^\circ}}\right)^\gamma\,
              T_{\rm SS}(\nv)\right]\left(\frac{\nu}{\nu_0}\right)^{\beta_s(\nv)},
            \end{equation}
            where $\gamma=1.5$, $T_{\rm LS}$ is the large-scale template at
            $\nu_0=23\,{\rm GHz}$, $T_{10^\circ}$ is a smoothed version of $T_{\rm LS}$ using
            a Gaussian kernel with FWHM $10^\circ$, $\beta_s(\nv)$ is the spectral index
            template and $T_{\rm SS}$ is the small-scale realization. Note that this
            procedure is different from the method used to generate the default sky
            templates provided with the {\tt PySM} package.
            
            For the polarized synchrotron templates {\tt PySM} follows a similar
            procedure to the intensity. To date, however, there is no precise determination
            of the spatial variation of the synchrotron spectral index, either in intensity or
            polarization (e.g. see \cite{2014ApJ...790..104F,2015arXiv150201588P}). Thus, for
            our purposes we use the K-band measurement of WMAP smoothed with a Gaussian kernel of
            $1^\circ$ FWHM as a large-scale amplitude template. {\tt PySM} then extrapolates it
            to different frequencies assuming the same spectral index template derived for
            intensity. This template is completed on small scales using a procedure similar
            to the one described above, where this time the amplitude of the small-scale
            component is modulated by the local mean polarized intensity $P=\sqrt{Q^2+U^2}$
            smoothed on a 10-degree scale. In this work we have not considered departures from
            a power-law spectral behaviour for synchrotron.
            \begin{figure}
              \centering
              \includegraphics[width=0.49\textwidth]{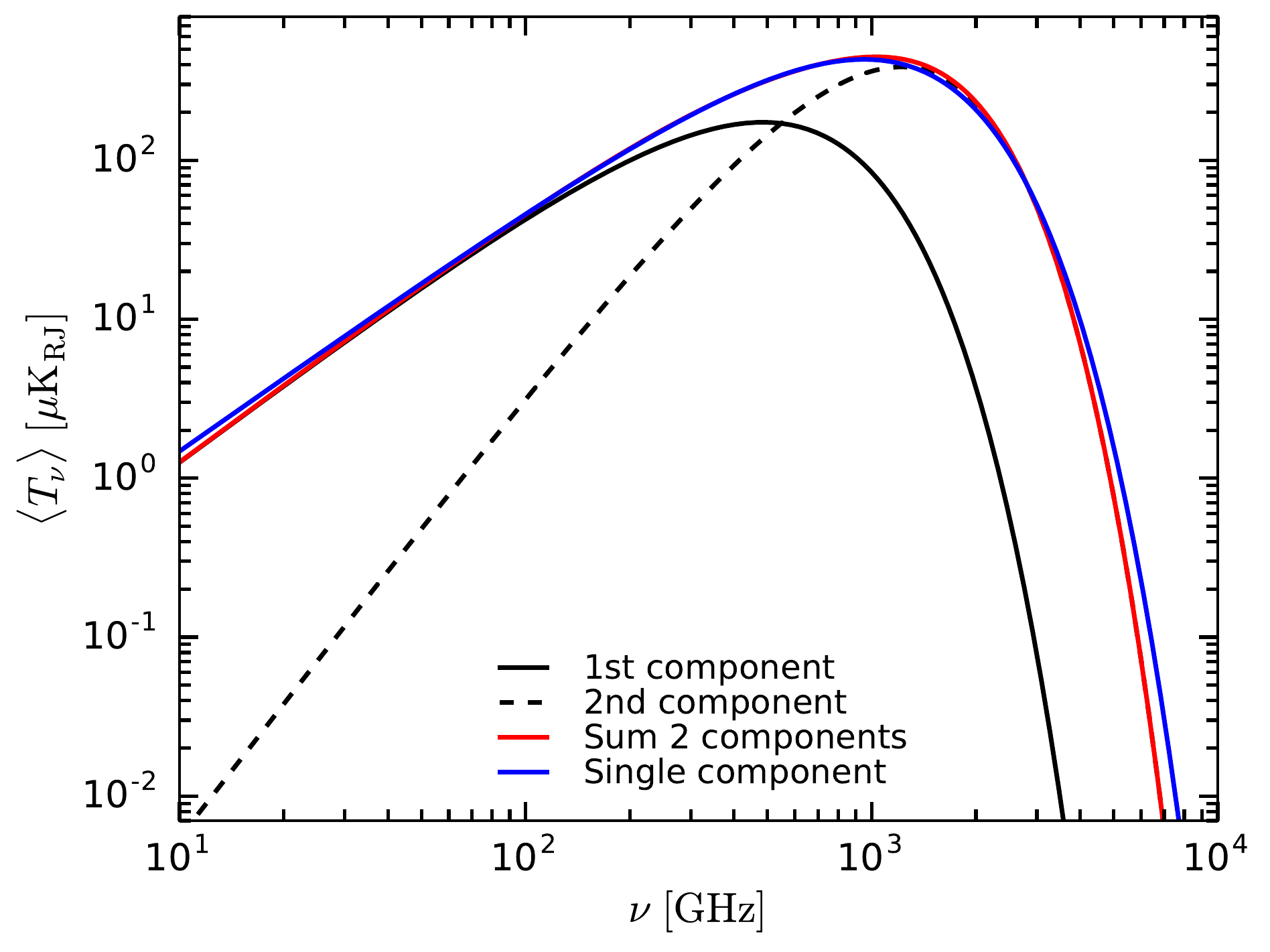}
              \caption{Frequency dependence of the mean dust temperature for the two-component
                       model of \cite{2015ApJ...798...88M} (red solid line, individual components
                       shown as black solid and dashed lines) and for the single MBB model of
                       \cite{2015arXiv150201588P} (blue solid line). Both models provide a good
                       fit to the data for most of the frequency range typically covered by
                       CMB experiments, and data at higher and lower frequencies would be
                       necessary to distinguish them.}\label{fig:freq_evol_dust}
            \end{figure}

      \item {\bf Thermal dust}: the long-wavelength tail of the thermal emission by
            Galactic dust grains heated by stellar radiation is the main foreground
            source at frequencies larger than $\sim100\,{\rm GHz}$. Furthermore, the
            alignment of non-spherical grains with the Galactic magnetic field
            generates linear polarization orthogonal to it and to the line of sight,
            thus making thermal dust arguably the most important contaminant for
            $B$-mode searches.
            
            The emission of thermal dust has been shown to be relatively well fit
            by a modified black-body spectrum (MBB) with a power-law emissivity
            \cite{2015arXiv150201588P}. In this work we have generated simulations
            of the thermal dust emission using a method similar to the one described
            above for synchrotron. As before, we use a template for the emission
            amplitude on large scales at a fixed frequency generated by {\tt PySM}, on top
            of which small-scale fluctuations are added as a high-pass filtered Gaussian
            realization of a power-law power spectrum
            $C_\ell\propto\ell^{\alpha_d}$, with $\alpha_d=-2.3$, modulated by the
            normalized local mean of the large-scale template smoothed on scales 
            $\theta_{\rm FWHM}=10^\circ$. This amplitude map is then extrapolated
            by {\tt PySM} to different frequencies using templates for the spectral
            parameters of the modified black-body intensity (spectral index
            $\beta_d$ and dust temperature $\Theta_d$). The explicit model used
            here is then:
            \begin{align}\nonumber
              T_{\rm dust}(\nu,\nv)=&
              \left[T_{\rm LS}(\nu_0,\nv)+
              \left(\frac{T_{10^\circ}(\nv)}{\bar{T}_{10^\circ}}\right)^\gamma\,
              T_{\rm SS}(\nv)\right]\times\\
              &\left(\frac{\nu}{\nu_0}\right)^{\beta_d(\nv)}
              \frac{B_\nu(\Theta_d(\nv))}{B_{\nu_0}(\Theta_d(\nv))},
            \end{align}
            where $\gamma=1.5$ and
            \begin{equation}
              B_\nu(\Theta)\equiv \frac{2h\nu^3}{c^2}
              \left[\exp\left(\frac{h\nu}{k\,\Theta}\right)-1\right]^{-1}
            \end{equation}
            is the black-body spectrum. We must note that, even though the amplitude
            of different foreground components should be spatially correlated, we
            have neglected this correlation on the small scales where we add power.
            The effect of this assumption should be irrelevant for the large-scale
            observables ($\ell\lesssim100$) we are interested in.
            
            For intensity {\tt PySM} uses the {\tt Commander} templates for the amplitude
            and spectral parameters at  $\nu_0=545\,{\rm GHz}$ \cite{2015arXiv150201588P}.
            In polarization, we use the {\tt Commander} templates for the $Q$ and $U$
            amplitudes at $\nu_0=353\,{\rm GHz}$, which are extrapolated to other
            frequencies using the same spectral parameter templates used for intensity.
            Note that this model is not completely realistic: the different alignment
            efficiency of different types of dust grains should induce a different frequency
            dependence in intensity and polarization, and there is evidence for this
            in the Planck data \cite{2015A&A...576A.107P} in terms of a global spectral index. 
            However, there is no estimate to date of the spatial variation of the
            polarized dust spectral index, and therefore we adopt the model above
            in order to simulate this spatial variation.
            
            It has been noted in the literature \cite{1999ApJ...524..867F,2015ApJ...798...88M}
            that a two-component dust model, with independent spectral indices and dust
            temperatures for both components, provides a marginally better fit when
            combining the Planck and DIRBE data. Although the joint emission from these
            two components can be qualitatively fit by a single MBB spectrum (see
            Fig. \ref{fig:freq_evol_dust}), future experiments might be sensitive to the
            differences between both models. Therefore we have run a number of
            simulations using a two-component dust model. For this, in intensity {\tt PySM}
            uses the amplitude and dust temperature templates provided by
            \cite{2015ApJ...798...88M}, together with their best-fit parameters. The
            corresponding polarization templates at 353 GHz are then generated as:
            \begin{equation}
             (Q,U)_{353}(\nv)=p_{353}(\nv)\,T_{353}^{2c}(\nv)\,
             (\sin2\gamma(\nv),\cos2\gamma(\nv)),
            \end{equation}
            where $T_{353}^{2c}$ is the predicted intensity of the two-component model
            at 353 GHz, and $p_{353}(\nv)$ and $\gamma(\nv)$ are maps of the polarized
            fraction and polarization angles at 353 GHz predicted by the best-fit
            {\tt Commander} single-component dust model.
            
            It is worth noting that, even though by using this second model we have tried
            to explore departures from our fiducial single-component model, the actual
            nature of thermal dust emission could be significantly more complicated than
            either of them.
            
      \item {\bf AME/Spinning dust.} The rotation (as opposed to vibration) of
            dust grains can also produce microwave emission, and this process is
            believed to be behind the so-called ``anomalous dust emission'' (AME),
            most prominent at low frequencies. Although the level to which this
            component is polarized is not clear, a failure to account for it could bias
            the measurement of $r$ in $B$-mode searches, and for this reason we have
            run a few simulations including this effect. 
            
            We use the AME templates provided in the {\tt PySM} package. In intensity,
            the code uses the best-fit {\tt Commander}  AME model and templates 
            \cite{2015arXiv150201588P}.  
            This model allows for two spinning dust components with different amplitudes
            and spectral parameters.  The model spectrum is computed using {\tt SpDust2}
            \cite{2009MNRAS.395.1055A, 2011MNRAS.411.2750S} for a cold neutral medium, and can 
            be rigidly shifted in $\log(\nu)$ by varying the peak frequency parameter. 
            The peak frequency of the first component varies spatially at degree scales in the range
            $\nu_{p1}(\nv) \sim 19 \pm 3$~GHz, but the second component is spatially constant at 
            $\nu_{p2} = 33$~GHz.  These spectra are then used to extrapolate two templates at 
            reference frequencies $\nu_1, \nu_2 = 22.8, 41.0$~GHz, which are also limited to
            degree-scale resolution. The model can be summarized as:
            \begin{equation}
            T^{\rm AME}_{\nu}(\nv)= T_{\nu_1}(\nv) f_{\rm {\tt SpDust}}(\nu_{p1}(\nv),\nu) +
            T_{\nu_2}(\nv) f_{\rm {\tt SpDust}}(\nu_{p2},\nu).
            \end{equation}
            
            The resulting total spectrum is much broader than those of the individual components, and 
            peaks in the range $\sim 20-30$~GHz.  It is stressed in \cite{2015arXiv150201588P} 
            that the second component is included only because a single component model left significant
            dust-correlated residuals.  The use of a second component is therefore not physically
            motivated, but is a convenient fit to the data. 

            In polarization, {\tt PySM} uses a simple model based on assuming a constant
            polarized fraction $p_{\rm AME}$ and using the polarization angle $\gamma(\nv)$ for
            thermal dust emission found at 353 GHz. Thus
            \begin{equation}
             (Q,U)_\nu(\nv)=p_{\rm AME}\,T^{\rm AME}_\nu(\nv)\,
             (\sin2\gamma(\nv),\cos2\gamma(\nv)).
            \end{equation}
            In our simulations we assumed a $2\%$ polarization fraction ($p_{\rm AME}=0.02$).
            Since there are physical reasons to expect that spinning dust should be almost
            unpolarized \cite{2016arXiv160506671D}, the model adopted here represents a
            conservative case. However note that other alternative models for AME, such as
            magnetic dust \cite{1999ApJ...512..740D} could be significantly more polarized.
            
    \end{enumerate}
    Table \ref{tab:sims} lists all the different simulations that were used for this
    work, corresponding to different variations of the models quoted above, together
    with different choices of the experiment design as well as the foreground cleaning
    algorithm. We note that the results quoted in this paper correspond to those extracted
    from a single simulation for each combination of experiment, sky and noise model, sky
    area and foreground cleaning method. However, we verified that our main results, in
    terms of the final uncertainty on the tensor-to-scalar ratio depend very little on the
    particular realization used.
    
  \subsection{Experimental setups} \label{ssec:method_exp}
    \begin{table*}
      \centering{
      \renewcommand*{\arraystretch}{1.6}
      \begin{tabular}{|c|c|c|}
        \hline
        Name    & Frequencies (GHz) & 
        RMS noise ($\mu{\rm K}_{\rm CMB}\,{\rm arcmin}/f_{\rm sky}^{1/2}$) \\
        \hline
        Stage-3 & (28, 41, 90, 150, 230, 353) & 
                  (171, 152, 14.2, 8.9, 16.5, $24^*$) \\
        Stage-4 & (30, 40, 85, 95, 145, 155, 215, 270) &
                  (29, 29, 4.7, 3.7, 3.5, 3.4, 5.2, 4.5)\\
        \hline
      \end{tabular}}
      \caption{Specifications for representative CMB experiments. For S3 we use the target
               frequencies and noise levels of AdvACT \cite{2015arXiv151002809H} and scale
               them up by a factor of $\sqrt{2}$ assuming that only half of the survey will
               be devoted to B-mode searches. For S4, \cite{Buza.inprep} choose a possible
               set of frequency channels in the atmospheric windows and noise levels designed
               to yield a map-level rms noise of $\sim1\uKam$ after foreground cleaning for
               $f_{\rm sky}=0.1$ \cite{Buza.inprep,S4.inprep}. Here the assumption is that
               the first three atmospheric windows would be covered by two different
               frequency channels. Note that in both cases the quoted noise levels are given
               in intensity, in antenna temperature units and for a full-sky experiment.
               When studying different sky areas the noise levels are therefore scaled
               with $\sqrt{f_{\rm sky}}$ accordingly. $^*$The only exception to this is
               the 353 GHz frequency channel for S3, which corresponds to the Planck 353
               GHz map, and therefore does not scale with sky area (see Section
               \ref{ssec:results_fisher} for further details). The noise levels above are
               quoted in intensity, and assume that the noise levels in $Q$ and $U$ will be
               a factor $\sqrt{2}$ larger.}\label{tab:cmbexp}
    \end{table*}
    The previous generation of ground-based CMB experiments, such as ACTPol
    \cite{2014JCAP...10..007N} and SPT-Pol \cite{2015ApJ...807..151K} have now been upgraded,
    or are being upgraded into so-called Stage-3 (S3) facilities, including AdvACT
    \cite{2015arXiv151002809H} and SPT-3G \cite{2014SPIE.9153E..1PB}. Looking ahead, S3
    experiments will be superseded by a Stage 4 (S4) experiment, likely to be
    built by combining the observing power of different ground-based facilities, with
    similar potential for wide sky coverage and significantly lower noise levels.
  
    Here we have considered two different experimental setups, corresponding to a Stage-3 
    AdvACT-like experiment, characterised by a $1.4$ arcmin. beam and
    $\sim8\,\mu K\,{\rm arcmin}$ rms noise (for a nominal $f_{\rm sky}=0.4$ sky coverage),
    and a future S4-like experiment, with roughly eight times lower noise. The frequency
    channels and noise levels used in both cases are summarized in Table \ref{tab:cmbexp}.
    For the S4-like experiment, the specifications are not yet well defined so we use map
    depths estimated by \cite{Buza.inprep}, derived by scaling the achieved performance of
    the BICEP2/Keck experiments \cite{2016PhRvL.116c1302B}.
    Note that, in the case of S3, we have also added the Planck 353 GHz data, which should
    help remove thermal dust. In the rest of this work, whenever we study the effects of
    using a reduced sky fraction, these noise levels were
    scaled down with $\sqrt{f_{\rm sky}}$ accordingly, assuming a constant observation time.
    In terms of defining the observable sky fraction, we will also assume that both
    experiments are located in the southern hemisphere.
        
    Each simulation consists of a set of $I,\,Q$ and $U$ maps in the frequency bands
    listed in Table \ref{tab:cmbexp}. These maps were generated using the HEALPix
    pixelization scheme \cite{2005ApJ...622..759G} with resolution parameter
    $N_{\rm side}=256$\footnote{Note that the pixel resolution used here is far worse than
    that allowed by both experiments.}, enough for detecting the large-scale $B$-mode
    signal, and instrumental Gaussian noise was added to every map.
    
    Atmospheric correlated noise is an important concern for ground-based experiments,
    in particular regarding large-scale observables such as primordial $B-$modes. For this
    reason, we have studied the effects of correlated noise by simulating the noise maps
    as Gaussian realizations of a power spectrum $N_\ell$ consisting of an uncorrelated and
    a correlated, power-law-like component:
    \begin{equation}
      N_\ell=\sigma_N^2\,\left[1+\left(\frac{\ell}{\ell_{\rm knee}}\right)^\gamma\right],
    \end{equation}
    where $\sigma_N^2$ is the noise variance per steradian (given by the white noise levels shown
    in Table \ref{tab:cmbexp}) and $\gamma=-1.9$. The parameter $\ell_{\rm knee}$ determines
    the scale below which the correlated component dominates, and we have studied the cases
    $\ell_{\rm knee}=0,\,50$ and 100, where the first case corresponds to purely white
    noise. By default we will report on the results with $\ell_{\rm knee}=0$, and we will
    analyze the effects of correlated noise in Section \ref{sssec:results_sims_corrnoise}.
    
    The value of the power-law index was chosen to roughly mimic the noise power spectrum
    achieved by the BICEP2/Keck experiments \cite{2016PhRvL.116c1302B}, and we assume the
    same shape for the noise power spectrum in intensity and polarization. The actual noise
    properties of specific experiments, however, will depend on a number of factors, such
    as atmospheric properties, instrumental specifications including modulation method, or
    survey strategy. We do not study these details further in this paper.

  \subsection{Map-based component separation}\label{ssec:method_bayes}
    In order to separate the different components that make up the total sky emission
    we have adopted here a map-based Bayesian scheme \cite{2004ApJS..155..227E,
    2008ApJ...676...10E,2009ApJ...701.1804D,2009MNRAS.392..216S,2016MNRAS.458.2032R}.
    An advantage of map-based methods over power-spectrum-based foreground cleaning (e.g.
    \cite{2015PhRvL.114j1301B}) is that the effect of spatially-varying foreground
    spectra can be taken into account, thus avoiding potential biases on
    large-scales. The advantage of Bayesian methods over blind methods such as the
    widely used internal linear combination method (ILC)
    \cite{2007ApJS..170..288H}, is that the resulting CMB maps are closer to optimal in
    terms of signal-to-noise ratio, and that the foreground uncertainties can be
    propagated consistently. On the other hand, the success of Bayesian methods relies
    on the goodness of the models adopted to describe the different components. We will
    discuss these differences further in Section \ref{sssec:results_sims_ilc}.
    
    \subsubsection{Description of the model}\label{sssec:method_bayes_model}
      In this method, the combined sky emission is modelled as a sum of several components with
      different frequency dependences. Let us consider a sky map with $N_\theta$ pixels,
      $N_p$ polarization channels (e.g. $T,\,Q,\,U$), and $N_\nu$ frequency bands. A model
      containing $N_c$ components can then be written, in general, as:
      \begin{equation}
        {\bf d}=\hat{F}\,{\bf T}+{\bf n},
      \end{equation}
      where
      \begin{itemize}
       \item ${\bf d}$ is the data, written as a vector of
             $N_\theta\times N_p\times N_\nu$ elements.
       \item ${\bf n}$ is the instrumental noise, also with
             $N_\theta\times N_p\times N_\nu$ elements.
       \item ${\bf T}$ is a vector with $N_\theta\times N_p\times N_c$ elements
             containing the amplitudes of the different components.
       \item $\hat{F}$ is a $(N_\theta\times N_p\times N_\nu)\times
             (N_\theta\times N_p\times N_c)$ matrix containing the spectral
             dependence of the different components, with the form:
             \begin{equation}
               F_{(i'\,p'\,\nu),(i\,p\,a)}\equiv
               f_a(\nu;{\bf b}_a^p(i))\,\delta_{i,i'}\delta_{p,p'}.
             \end{equation}
             Here $i,\,p,\,\nu$ and $a$ label the different angular pixels,
             polarization channels, frequency bands and sky components respectively,
             and ${\bf b}$ is a set of spectral parameters. In our case these are
             the synchrotron spectral index and the dust temperature and spectral
             index. Explicitly, for the three components considered here, $\hat{F}$ is
             \begin{align}
               {\rm CMB}&:\,f_C(\nu)=\left.
                \frac{dB_\nu(\Theta)}{d\Theta}\right|_{\Theta=\Theta_{\rm CMB}},\\
               {\rm synch.}&:\,f_s(\nu;\beta_s)=
                \left(\frac{\nu}{\nu_0^s}\right)^{\beta_s}\\
               {\rm dust.}&:\,f_d(\nu;\beta_d,\Theta_d)=\left(\frac{\nu}{\nu_0^d}\right)^{\beta_d}
                \frac{B_\nu(\Theta_d)}{B_{\nu_0^d}(\Theta_d)}.
             \end{align}
      \end{itemize}
      
      The method then consists of sampling the distribution of the free parameters of the model,
      which are given by:
      \begin{itemize}
       \item {\bf Amplitudes}, ${\bf T}$. $3\times N_p\times N_{\theta}$ of them.
       \item {\bf Spectral indices}, ${\bf b}^p_a$. We will assume independent spectral
             indices in intensity and polarization, but a common index for $Q$ and $U$.
             Furthermore, we will make the simplifying assumption that spectral indices
             vary only over pixels larger than our
             base resolution. Labelling the total number of such pixels as
             $N'_\theta<N_\theta$ we then have $2\times3\times N'_\theta$ spectral
             parameters.
      \end{itemize}
      We can compare the total number of parameters of the model,
      $N_{\rm par}=3\times N_p\times N_\theta+6\times N'_\theta$, with the total number of data
      points, $N_{\rm data}=N_p\times N_\nu\times N_\theta$, to see that we need
      $N_\nu>3$ frequency channels to prevent the system from becoming overparametrized.

      Using Bayes' Theorem, we can write the posterior distribution for the model
      parameters as
      \begin{equation}
        p({\bf T},{\bf b}|{\bf d})\propto p_l({\bf d}|{\bf T},{\bf b})\,p_p({\bf T},{\bf b}),
      \end{equation}
      where $p_p$ is the prior distribution for the parameters (which we will discuss
      later on) and $p_l$ is the Gaussian likelihood, given by
      \begin{equation}\label{eq:like1}
        -2\log p_l({\bf d}|{\bf T},{\bf b})=C+\left[{\bf d}-\hat{F}\,{\bf T}\right]^T
        \hat{N}^{-1}\left[{\bf d}-\hat{F}\,{\bf T}\right].
      \end{equation}
      Here $\hat{N}\equiv\langle{\bf n}\,{\bf n}^T\rangle$ is the noise covariance
      matrix, which we assume to be uncorrelated between frequency and
      polarization channels 
      \begin{equation}
        N_{(i\,p\,\nu),(i'\,p'\,\nu')}=N_{i,i'}^{p\nu}\delta_{pp'}\delta_{\nu\nu'}.
      \end{equation}
      In reality, instrumental and atmospheric effects will induce correlations between
      these channels.
      Here we will further assume, for simplicity, that the noise covariance is white,
      so that
      \begin{equation}
        (\hat{N}^{-1})_{(i\,p\,\nu),(i'\,p'\,\nu')}=\sigma_{\nu,p}^{-2}
        \delta_{ii'}\delta_{pp'}\delta_{\nu\nu'},
      \end{equation}
      where $\sigma_{\nu,p}^2$ is the per-pixel noise variance.

      Note that, by assuming uncorrelated noise, the posterior distribution can be
      written as a product of distributions for the individual large pixels over
      which the spectral indices are allowed to vary. This greatly simplifies the task
      of sampling the amplitude and spectral parameters, but will in general not be
      a valid assumption in the presence of atmospheric noise for ground-based
      experiments. Nevertheless, even in the presence of spatially correlated noise,
      which we study in the subsequent sections, this assumption should not introduce
      a bias in the final component-separated maps as long as the different frequency
      channels are appropriately weighted according to their overall noise variance.
      The variance of the output maps, however, will be sub-optimal; in the analysis
      of real data the noise correlations would be included.

      Note also that the logarithm of the posterior is a quadratic function of the
      amplitudes ${\bf T}$ (assuming that the prior on ${\bf T}$ takes a Gaussian form).
      We take advantage of this property implementing two different methods to carry
      out the sampling. First, the amplitudes can be directly sampled (with a $100\%$
      acceptance rate) as Gaussian random fields separately from the spectral index
      using Gibbs sampling methods as in e.g. \cite{2008ApJ...676...10E,
      2009ApJ...701.1804D,2011MNRAS.418.1498A}. Secondly, it is possible
      to marginalize over the amplitudes analytically, and thus sample only the spectral
      indices from their marginal distribution. The latter method significantly
      improves the performance of the algorithm. We give further details about the
      advantages and implementation of these methods in Appendix \ref{app:methods}.

    \subsubsection{Priors}\label{sssec:method_bayes_priors}
      In this work we imposed loose Gaussian priors on the spectral parameters,
      with $\beta_s=-3\pm0.5$, $\beta_d=1.54\pm0.5$ and $\Theta_d=20\pm0.5\,{\rm K}$.
      The width of these priors is large enough to avoid biases in the final maps,
      and we verified that they seldom drive the posterior distribution.
      Besides these, we included a ``volume prior'' designed to take into account the
      volume of likelihood space for non-linear parameters. This is described in detail
      in Appendix \ref{app:volume}, and is equivalent to the widely used Jeffreys prior for
      the spectral parameters \cite{2008ApJ...676...10E}.

  \subsection{Measuring $r$}\label{ssec:method_measure}
    After foreground cleaning we are left with a map of the mean and variance of the
    CMB intensity and polarization with fully propagated foreground uncertainties.
    From this map we determine $r$ and its uncertainty using a power-spectrum based
    likelihood, assuming that the $BB$ bandpowers are Gaussianly distributed:
    \begin{widetext}
    \begin{equation}
      -2\ln{\cal L}(r,A_L)={\rm const.}+
      \sum_{k,k'}\left[\hat{B}_k-B^{\rm th}_k(r,A_L)\right]
      \left(\Sigma^{-1}\right)^{-1}_{kk'}
      \left[\hat{B}_{k'}-B^{\rm th}_{k'}(r,A_L)\right],
    \end{equation}
    \end{widetext}
    where $\hat{B}_k\equiv\sum_{\ell}W^k_\ell \hat{C}^{\rm BB}_\ell$ and $B^{\rm th}_k$ are
    the measured and model $B$-mode bandpowers, and $\Sigma_{kk'}$ is the covariance matrix
    of $\hat{B}_k$. We model the $B$-mode power spectrum in terms of a primordial and a
    lensing component, each multiplied by a free amplitude parameter ($r$ and $A_L$
    respectively):
    \begin{equation}
      C^{\rm th}_\ell=r\,C^{\rm prim}_\ell(r=1)+A_L\,C^{\rm lens}_\ell.
    \end{equation}
    Here the primordial and lensing templates are held fixed to fiducial $\Lambda$CDM
    values.

    \begin{figure*}
      \centering
      \includegraphics[width=0.49\textwidth]{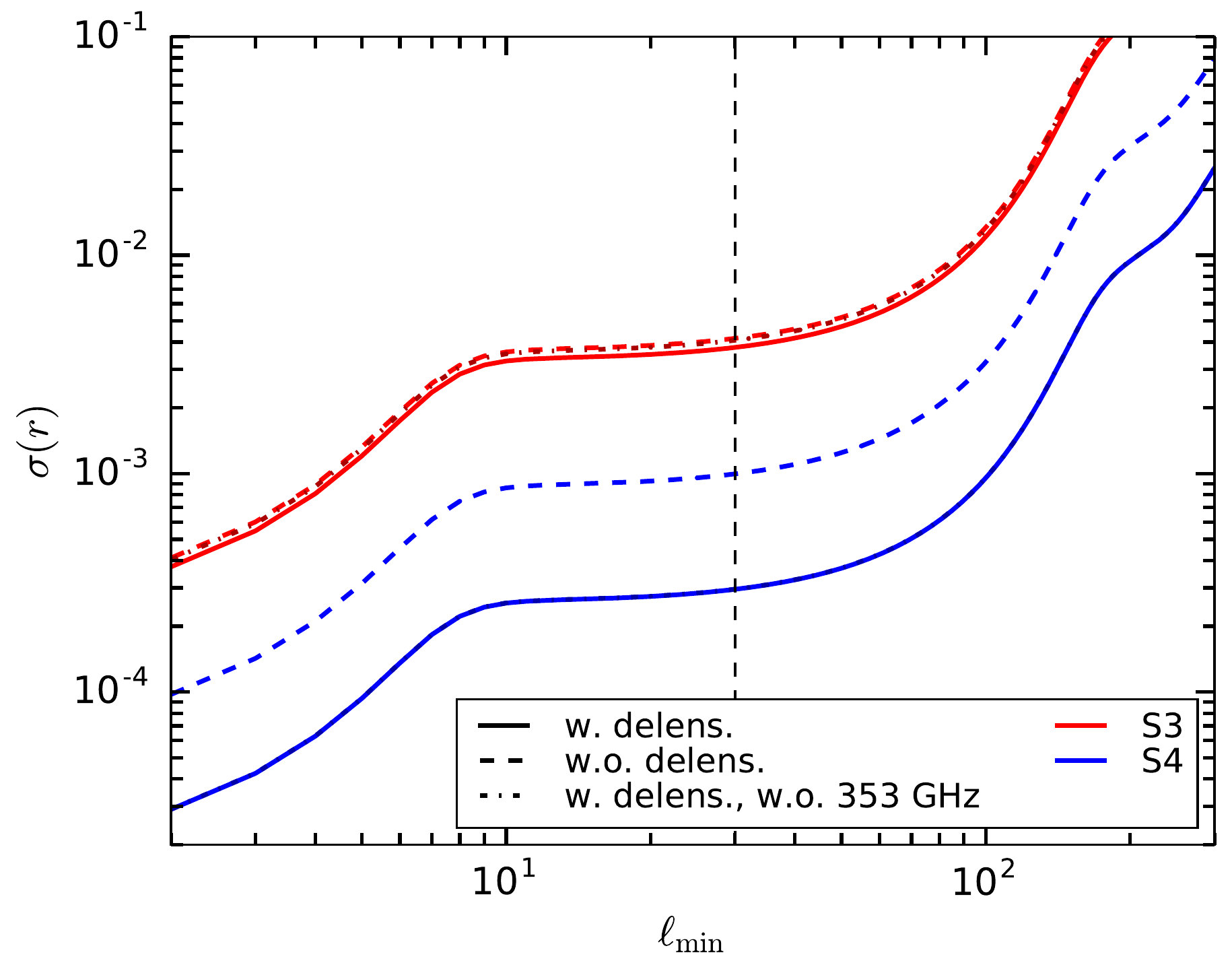}
      \includegraphics[width=0.49\textwidth]{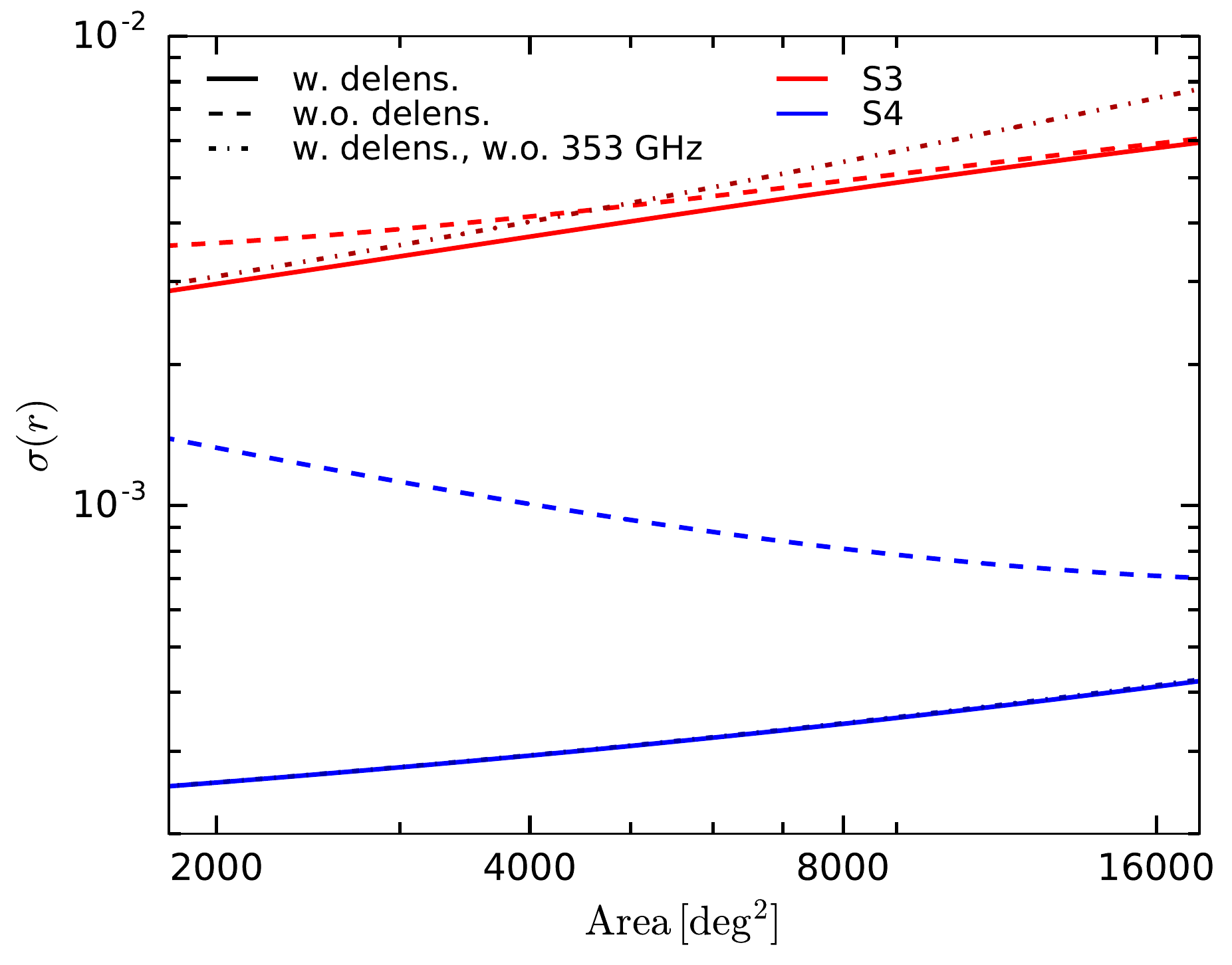}
      \caption{{\sl Left panel:} 68\% uncertainty levels on $r$ as a function of the minimum
               multipole included in the analysis for a fiducial sky area of
               $4000\,{\rm deg}^2$. Red and blue lines correspond to S3 and S4
               respectively. Solid (dashed) lines show the results with (without)
               delensing, and the dot-dashed lines correspond to the same experiments
               after excluding the Planck 353 GHz channel.
               {\sl Right:} uncertainty on $r$ as a function of sky area for a fixed
               observation time and for a fiducial $\ell_{\rm min}=30$. The Figure
               uses the same color code and line styles used in the left panel.
               Note that, while the Planck 353 GHz channel could help reduce the final
               uncertainty on $r$ for S3, especially for larger sky areas (higher noise),
               it is irrelevant for S4, given its lower noise levels (the blue solid and
               dot-dashed lines are indistinguishable).}
      \label{fig:fisher}
    \end{figure*}
    Note that the power spectrum bandpowers are not Gaussianly distributed. However,
    at sufficiently large $\ell$, the central limit theorem guarantees that their
    distribution can be well approximated as such, since they are determined by
    averaging over all $|a_{\ell m}|$'s corresponding to the same $\ell$. Since
    many ground-based experiments are expected to be limited by atmosphere-related
    systematic effects on scales $\ell\lesssim30$, using this approximation is justified.
    
    Since ground-based observations from a single site can not fully cover the
    celestial sphere and, in any case, Galactic foregrounds prevent a clean
    measurement of primordial $B$-modes on the full sky, the angular power spectrum
    must be computed in the presence of a sky mask. There are several approaches to
    this problem in the literature, which range from the optimal approaches of the
    maximum likelihood estimator or the minimum-variance quadratic estimator
    \cite{1997PhRvD..55.5895T} to the minimal approach of pseudo-$C_\ell$ estimators
    \cite{2002ApJ...567....2H}. The latter approach should be only marginally
    non-optimal for simple masks and non-steep power spectra (which is the case for
    $C_\ell^{BB}$), and therefore has been our method of choice for this work.
    
    However, the use of pseudo-$C_\ell$ codes for polarized signals (or, in general,
    any spin-2 field) is complicated by the fact that a straightforward implementation
    of the method will give rise to a non-negligible contamination of $E$ into $B$
    in the variance of the estimator. Since the CMB $E$-mode signal is much larger
    than the $B$-modes, this effect can make the pseudo-$C_\ell$ estimator severely
    sub-optimal. This problem can be solved by designing a pure-$B$ pseudo-$C_\ell$
    estimator \cite{2006NewAR..50.1025S}, which requires a non-trivial apodization
    around the edges of the mask. Since the aim of this paper is to assess the effect
    of foregrounds on the measurement of primordial $B$-modes, rather than the
    complications of power-spectrum estimation, we sidestep this problem by taking
    the following steps in each simulation:
    \begin{enumerate}
      \item Clean the foregrounds on the full sky.
      \item Rotate the foreground-cleaned maps from $(T,Q,U)$ into the
            (pseudo) scalars $(T,E,B)$ on the full sky.
      \item Apply the mask on the full-sky (pseudo)scalar maps.
      \item Run a spin-0 pseudo-$C_\ell$ algorithm on the masked $(T,E,B)$ maps.
    \end{enumerate}
    Thus this process preserves the complications of cut-sky observations (increased
    sample variance, correlations between band powers) while mimicking an optimal
    measurement of the $B$-mode power spectrum. We note that this method will yield
    smaller error bars than are actually achievable in a realistic situation. We can
    estimate the magnitude of this effect by noting that, as reported in
    \cite{2013PhRvD..88b3524F}, the pure-$B$ pseudo-$C_\ell$ estimator yields errors
    that are at most a factor of $\sim2$ larger than the theoretical $\propto
    f^{-1}_{\rm sky}$ expectation, while we estimate the standard deviation of
    our spin-0 pseudo-$C_\ell$ to be a factor of $\sim1.3$ larger than this ideal
    case. Thus, the uncertainties in the $B$-mode power spectrum in a realistic
    scenario would be, assuming a sub-optimal pseudo-$C_\ell$ approach, at most
    a factor $\sim1.5$ larger than those reported here.
    
    The details of the pseudo-$C_\ell$ method have been widely described in the
    literature, and we only quote the main details here. We compute the $BB$ power
    spectrum in bandpowers, estimated from the power spectrum of the cut-sky
    anisotropies as
    \begin{equation}
      \hat{B}_k=\sum_{k'}(\hat{M}^{-1})_{kk'}\,\tilde{B}_{k'},
    \end{equation}
    where
    \begin{equation}
      \tilde{B}_k=\sum_\ell\frac{W^k_\ell}{2\ell+1}\sum_m |\tilde{B}_{\ell m}|^2,
    \end{equation}
    $\tilde{B}_{\ell m}$ are the spherical harmonic coefficients of the masked
    $B$-mode map and $\hat{M}$ is the cut-sky coupling matrix. The latter depends only
    on the mask applied to the data, and its analytic expression can be found in
    \cite{2002ApJ...567....2H}. For this work we have used top-hat bandpowers
    characterised by a width $\Delta\ell$:
    \begin{equation}
      W^k_\ell=\frac{1}{\Delta\ell}\Theta(\ell-\ell_k)\Theta(\ell_k+\Delta\ell-\ell),
    \end{equation}
    where $\Theta(x)$ is the Heavyside function.
    
    We avoid the problem of noise bias by using only cross-correlations between
    simulations run with the same CMB signal but different noise realizations.
    This mimics the usual approach of cross correlating splits of the full data
    in CMB experiments. Finally, for each simulation we compute the covariance matrix
    of the bandpowers $\Sigma_{k,k'}$ from 1000 Gaussian realizations of the
    signal and noise $BB$ power spectrum measured from the two simulations. These
    realizations were cut using the same mask used in the analysis of the simulations,
    and therefore we fully account for possible non-zero correlations between bandpowers.

\section{Results}\label{sec:results}
      \begin{figure}
        \centering
        \includegraphics[width=0.49\textwidth]{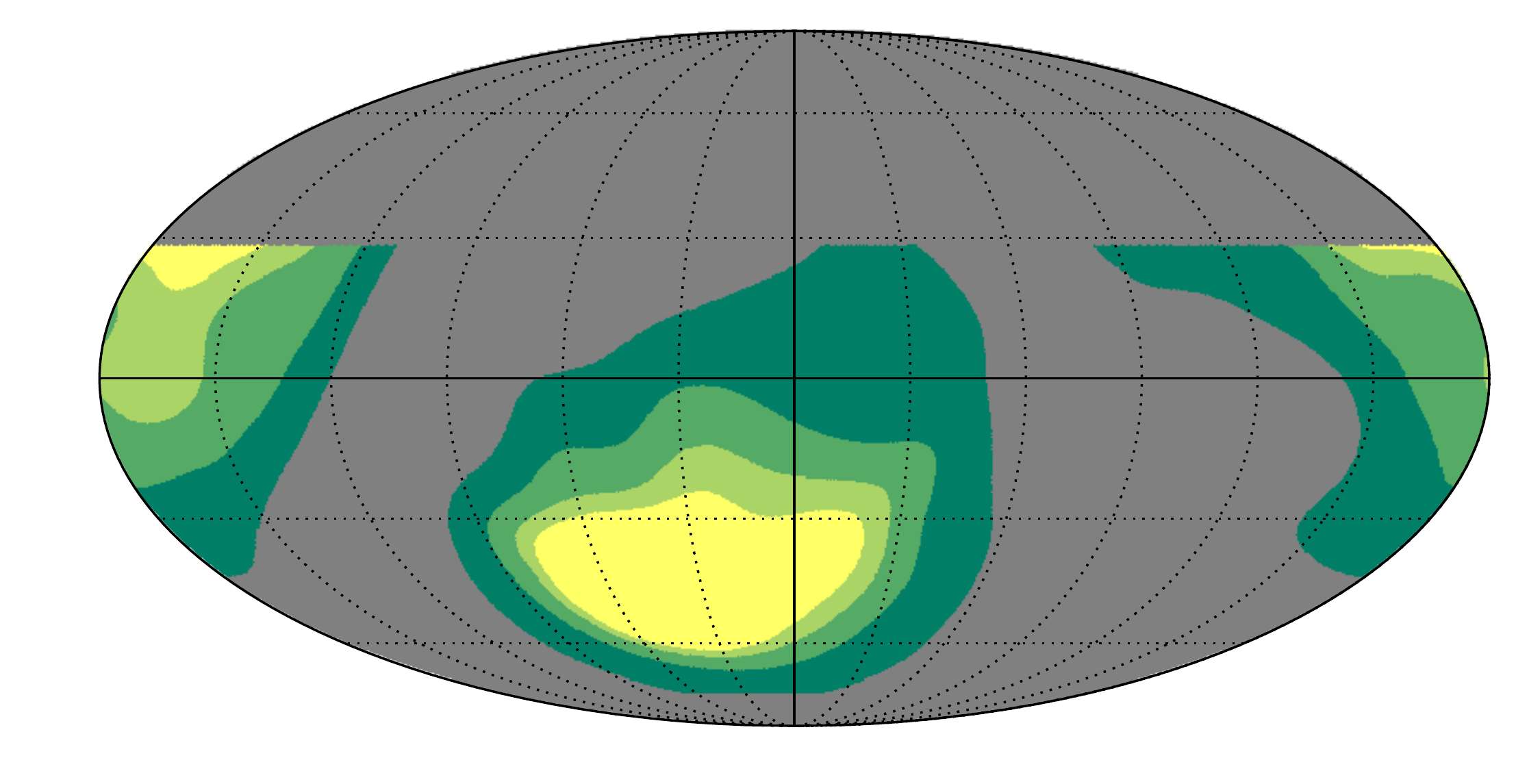}
        \caption{Sky masks used in the analysis, corresponding to the cleanest 2000,
                 4000, 8000 and 16000 ${\rm deg}^2$ of the sky accessible from Chile
                 in terms of foreground contamination.}
        \label{fig:masks}
      \end{figure}
  \subsection{Fisher matrix forecasts}\label{ssec:results_fisher}
    As a preliminary step, and in order to have an estimate of the most optimistic
    constraints on $r$ one can expect from our two model experiments, we have computed
    their corresponding Fisher forecast uncertainties. For this we assume global foreground
    spectral parameters $\beta_s=-3,\,\beta_d=1.54\,$ and $\Theta_d=20.9\,{\rm K}$, and a
    fiducial value of $r=0$. The foreground spectral parameters were held fixed, and thus
    these forecasts will yield the best possible uncertainties on $r$. Moreover, we assume
    a delensing factor $f_{\rm dl}$ related to the map noise level as described in 
    \cite{2016JCAP...03..052E}.
    
    The left panel of Figure \ref{fig:fisher} shows the expected 1$\sigma$ constraints as
    a function of the minimum multipole $\ell_{\rm min}$ included in the analysis with and
    without delensing (solid and dashed lines respectively) for S3 and S4 (red and blue
    respectively). The results in the absence of the Planck 353 GHz map are shown as dot-dashed
    lines in both cases, and all curves assume a sky fraction $f_{\rm sky}=0.1$ for both
    experiments. As is evident from the Figure, while delensing is vital for S4 in order to
    significantly reduce the uncertainties on $r$, it is a lot less relevant for S3.
    Furthermore, the contribution of the 353 GHz map is irrelevant for S4, while it has
    a non-negligible impact for S3.
     \begin{figure*}
       \centering
       \includegraphics[width=0.49\textwidth]{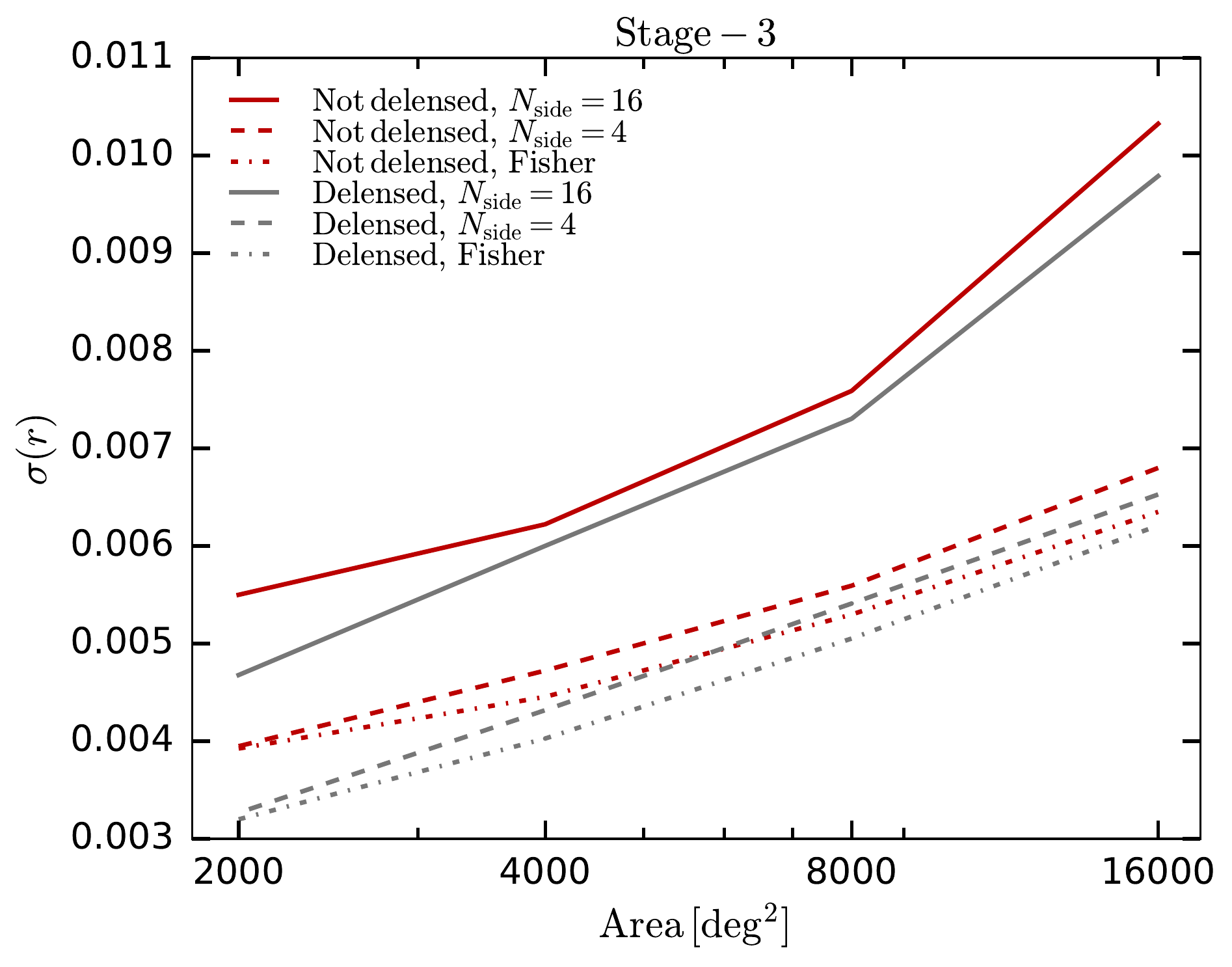}
       \includegraphics[width=0.49\textwidth]{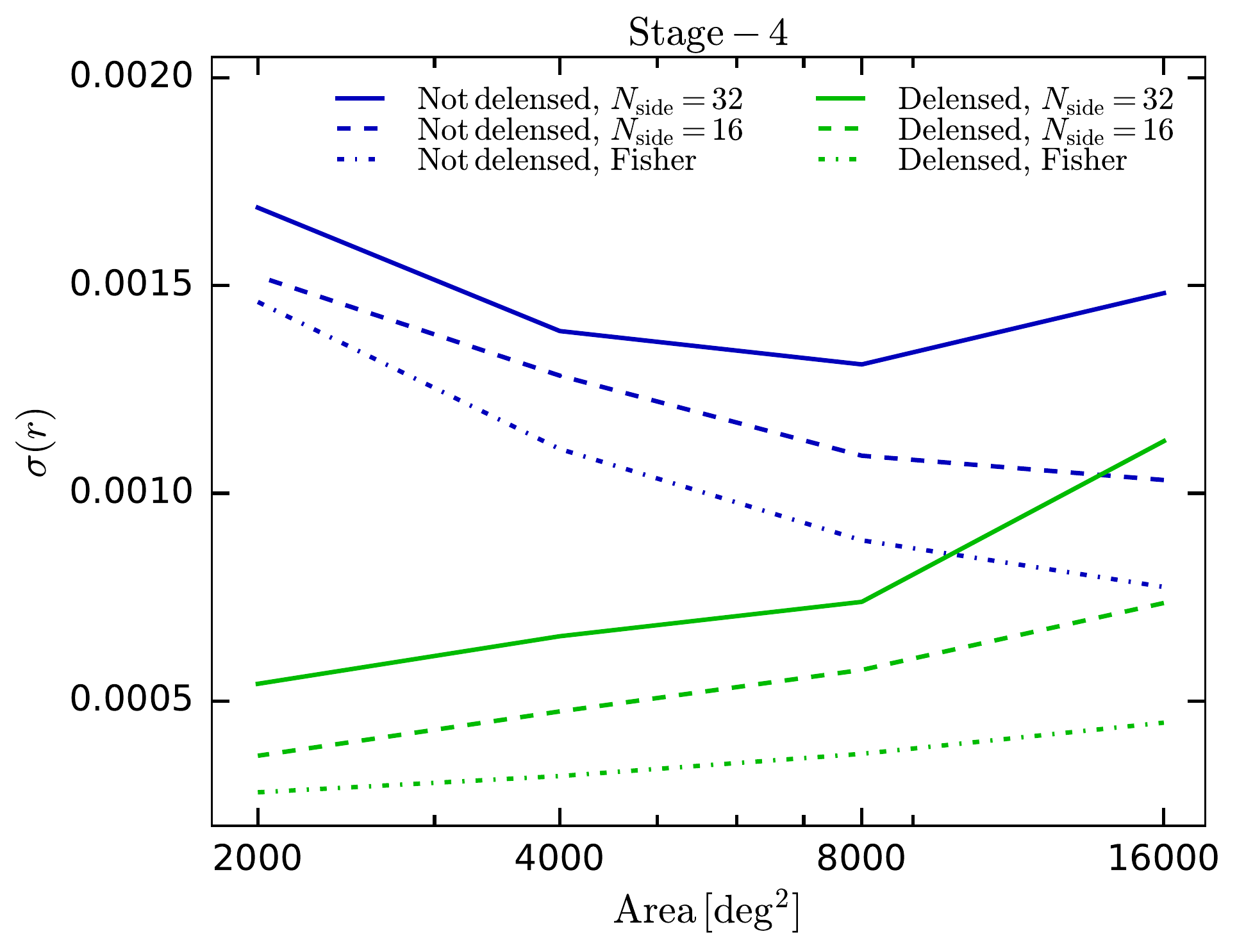}
       \caption{Simulated 68\%-level uncertainties on $r$. {\sl Left}: results for S3 
                with (without) delensing in gray (red). Results are shown for foregrounds
                cleaned assuming constant spectral indices on pixels of resolution
                $N_{\rm side}=4$ and 16 (solid and dashed lines respectively).
                {\sl Right}: results for S4 with (without) delensing in green (blue), and
                for spectral indices assumed constant on pixels of size $N_{\rm side}=16$
                and 32 (solid and dashed lines respectively). In both panels, the dot-dashed
                lines show the Fisher matrix forecasts, which assume fixed spectral indices.}
       \label{fig:sr_fiducial}
     \end{figure*}
    
    Although the high-signal region $\ell_{\rm min}\lesssim10$ caused by the reionization
    bump is likely inaccessible for ground-based experiments, the plateau between
    $\ell_{\rm min}\sim10$ and $\ell_{\rm min}\sim70$ can still
    be used to impose competitive constraints on $r$. Beyond $\ell_{\rm min}\sim70$, the
    sensitivity to $r$ decreases sharply, and therefore it is important to cover
    the aforementioned plateau. In order to ensure that these large scales are
    sufficiently well sampled, we will only consider sky fractions
    $f_{\rm sky}\gtrsim0.05$ for the rest of the analysis. The right panel of Fig.
    \ref{fig:fisher} shows the dependence of $\sigma(r)$ on $f_{\rm sky}$ for our
    fiducial value of $\ell_{\rm min}=30$.

  \subsection{Simulated forecasts} \label{ssec:results_sim}
    \subsubsection{Foreground masks} \label{sssec:results_sim_masks}
      In order to study the effects of foregrounds on $B$-mode searches as a function of sky
      fraction we have designed sky masks covering the cleanest 16000, 8000, 4000, and 2000
      square degrees of the southern sky. We do so by first creating a map of the combined
      foreground emission by synchrotron and dust at 100 GHz smoothed on scales of
      $\sim20^\circ$ and then selecting the connected regions in this map with the lowest
      foreground emission in $P\equiv\sqrt{Q^2+U^2}$. We further restrict these regions to
      lie in the range of declination ${\rm dec}\in[-75^\circ,28^\circ]$. The resulting
      masks are shown in Fig. \ref{fig:masks}.

   \subsubsection{Results: fiducial foregrounds} \label{sssec:results_sims_fid}
     We start by examining the fiducial simulations, with a single thermal dust component and
     without polarized AME. One of the free parameters of our method is the size of the large
     pixels, over which the spectral parameters are assumed to be constant. Smaller pixels
     allow us to capture the spatial variation of spectral indices more faithfully, at the
     cost of including a larger number of model parameters, which inevitably increases the
     final map-level noise (and consequently the uncertainty on $r$). We study the minimum
     resolution (in terms of the HEALPix $N_{\rm side}$ resolution parameter) needed to avoid
     biasing our $r$ measurement for the Stage-3 and Stage-4 experiments, given the fiducial
     foregrounds model adopted here. Figure \ref{fig:sr_fiducial} shows the $68\%$ uncertainty
     on $r$ for different resolutions for S3 and S4 (left and right panels respectively). The
     increase in $\sigma(r)$ caused by using more finely resolved spectral parameters is
     evident, and can be as large as a factor of $\sim1.8$ for S3. We determined that
     $N_{\rm side}=4$, corresponding to an angular scale of $\sim14^\circ$ is enough to
     avoid a bias in $r$ at the $1.5\sigma$ level (i.e. $\bar{r}<1.5\sigma(r)$) given the
     noise levels of S3, while at least $N_{\rm side}=16$ ($\sim3.5^\circ$)
     was needed for S4 following the same criteria. Note that the resolution needed to fit
     for spectral parameters depends directly on the properties of the foregrounds, and 
     therefore the values quoted here are specific to the simulations described in Section
     \ref{ssec:method_sims}.
     
     Figure \ref{fig:sr_fiducial} also shows the effect of delensing in the final uncertainties.
     As we saw before, the improvement for S3 is only moderate, while the noise level of S4
     makes the measurement of $r$ cosmic-variance limited in the absence of delensing. This
     can be clearly seen in the right panel of Fig. \ref{fig:sr_fiducial}, not just as an
     increase in $\sigma(r)$ with respect to the delensed case, but also in the fact that,
     without delensing, a larger area is always preferred, in spite of its higher noise level.
     This trend reverts after delensing, when the measurement of $B$-modes becomes again
     noise-dominated. This is consistent with findings in e.g.
     \cite{2016JCAP...03..052E,Buza.inprep}.
     \begin{figure*}
       \centering
       \includegraphics[width=0.49\textwidth]{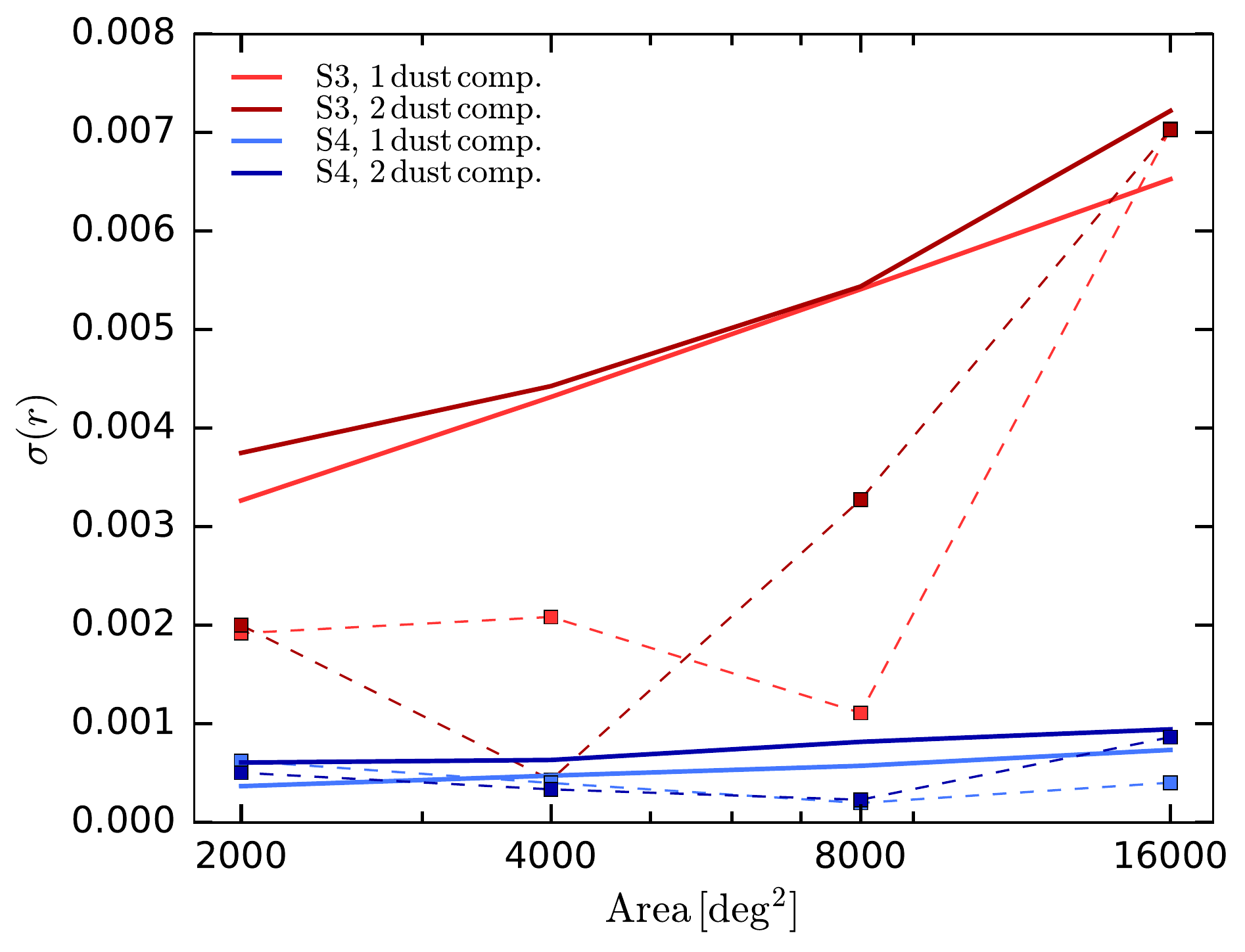}
       \includegraphics[width=0.49\textwidth]{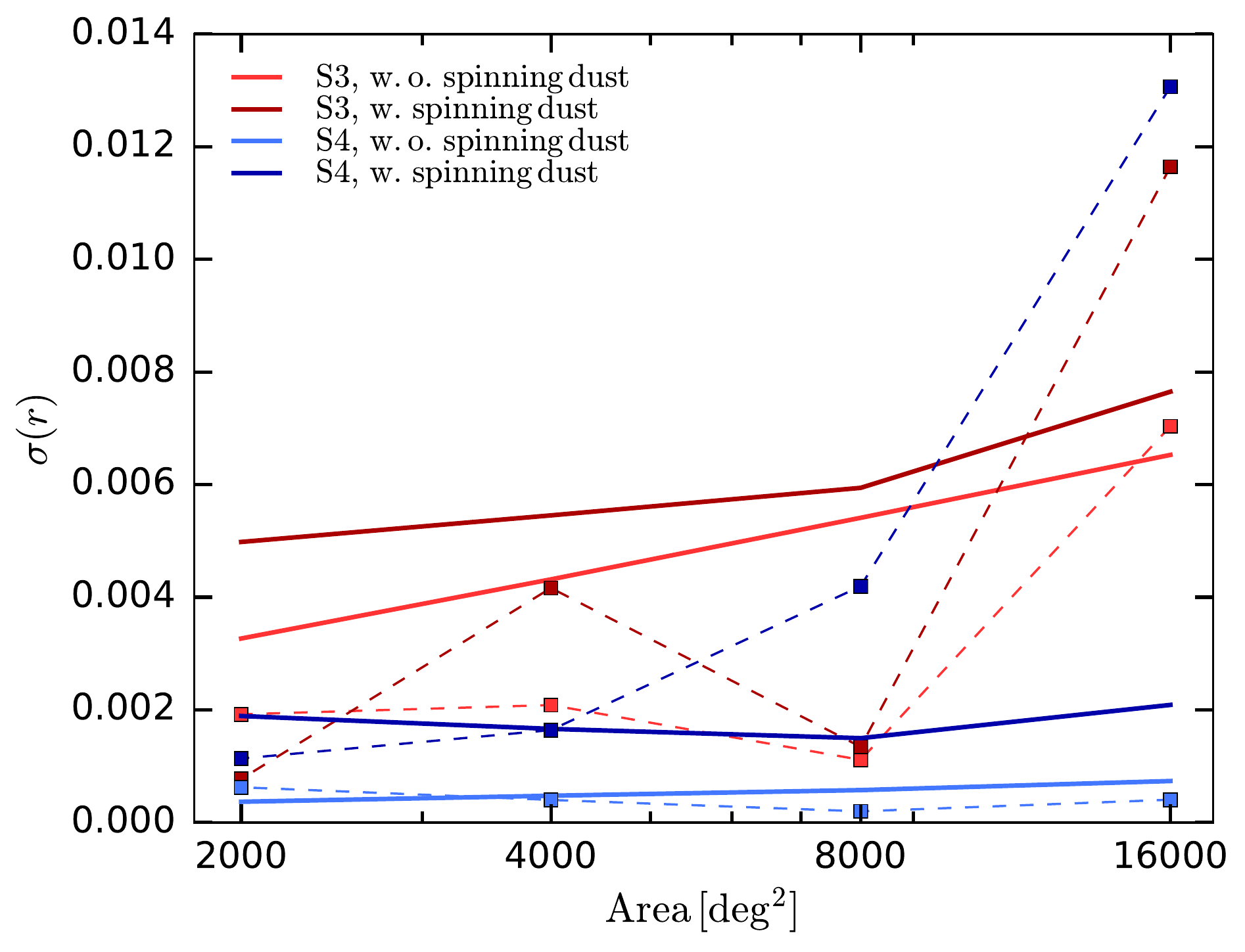}
       \caption{Simulated 68\%-level uncertainties (solid lines) and measured value (dashed lines)
                of $r$ for foreground simulations that depart from our fiducial foreground model
                (used by the component separation code). Results are shown for S3 (red) and S4
                (blue) for the fiducial and alternative foreground simulations (light and dark
                colors respectively). {\sl Left:} results for 2 versus 1 independent thermal dust
                components. In this case we do not observe a significant ($>1.5\sigma$) bias on
                $r$. {\sl Right:} results with and without $2\%$-polarized AME. The presence
                of this component would significantly bias the best-fit value of $r$ in
                the case of S4 if neglected at the component-separation stage, particularly for
                larger (more contaminated) sky areas.}
       \label{fig:sr_deviations}
     \end{figure*}
     
     Finally, Figure \ref{fig:sr_fiducial} also shows the value of $\sigma(r)$ predicted by
     the Fisher matrix approach described in the previous sections (which assumes fixed
     foreground spectral indices). The Fisher prediction is always more optimistic than our
     simulated results, although both are similar for S3 in the $N_{\rm side}=4$ case. This
     makes sense as, in the absence of spatially varying spectral parameters (i.e. in the
     limit $N_{\rm side}\rightarrow0$), the Fisher prediction should be recovered.
     
     In what follows, all results will be presented in the delensed case and for spectral
     parameters sampled in pixels of resolution $N_{\rm side}=4$ and $N_{\rm side}=16$ for
     S3 and S4 respectively.
     
   \subsubsection{Results: deviations from the fiducial model} \label{sssec:results_sims_alt}
    \begin{table}
      \centering{
      \renewcommand*{\arraystretch}{1.6}
      \begin{tabular}{|c|c|c|c|c|}
        \hline
        \multirow{2}{*}{Area (deg$^2$)} & \multicolumn{2}{c|}{$\chi^2/{\rm dof}-1$} &
        \multicolumn{2}{c|}{${\rm PTE}(\chi^2)$} \\
        \cline{2-5}
                       & No AME & W. AME & No AME & W. AME \\
        \hline
         2000          &  $-5.91\times10^{-4}$ & $2.81\times10^{-3}$ & 0.24 & $0.099$ \\
         4000          &  $2.54\times10^{-4}$  & $2.31\times10^{-3}$ & 0.18 & $0.072$ \\
         8000          &  $1.02\times10^{-3}$  & $2.52\times10^{-3}$ & 0.44 & $0.013$ \\
        16000          &  $5.62\times10^{-4}$  & $1.93\times10^{-3}$ & 0.61 & $0.008$ \\
        \hline
      \end{tabular}}
      \caption{Values of the map $\chi^2$ for the mean amplitudes and spectral indices for
               S4 simulations with and without a polarized AME component, as well as the
               associated $p$-values. Although the overall $\chi^2$ per degree of freedom
               is close to 1 in all cases, the unaccounted-for AME component is associated
               with much smaller PTEs, which in the case of the higher sky areas would be
               a clear sign for missing components in the model used by the foreground
               cleaning algorithm. In this case, all spectral parameters were allowed
               to vary in pixels of resolution $N_{\rm side}=16$.}
      \label{tab:chi2}
    \end{table}
     One of the drawbacks of Bayesian component separation methods is that specific models
     have to be assumed regarding the properties of the different components (e.g.
     frequency or spatial dependence). An incorrect modelling can therefore introduce biases
     that could potentially leak into the final cosmological parameters. In order to explore
     this possibility we have repeated the exercise described in the previous section on
     simulations containing foregrounds that are not described by the model used by our
     component separation code (i.e. single thermal dust component and power-law
     synchrotron). Specifically we generated simulations containing two thermal dust
     components, described by the model of \cite{2015ApJ...798...88M}, and containing a
     $2\%$ polarized AME component.
     
     The results are summarized in Figure \ref{fig:sr_deviations}, which shows the uncertainty
     on $r$ as a function of sky area as solid lines, together with its best-fit value as
     dashed lines. The left panel shows the results for the simulations with 2 dust components.
     Both for S3 and S4, the single-component model is able to fit well the joint emission of
     the 2 dust components, and no significant bias on $r$ is observed. The slight increase
     in $\sigma(r)$ with respect to the single-component model is caused by the different
     weighting of the different frequency maps needed to accommodate the spectral behavior of
     the 2-component model. We therefore conclude that the existence of a second dust component
     (with spectral properties compatible with current data) would not generate an important
     foreground bias in the measurement of primordial B-modes if unaccounted for.
     
     The right panel of Fig. \ref{fig:sr_deviations} shows the results in the presence of
     AME. As before, the bias on $r$ induced by the unaccounted-for AME is masked
     by the large noise levels of S3, and no significant bias can be appreciated. In the case
     of S4, however, we observe a dramatic biasing of $r$ caused by the presence of polarized
     AME, which lies above the $2\sigma$ level for the largest sky areas. Given such
     a large foreground bias, it is worth exploring whether the existence of an unmodelled
     component would have been detected at the foreground-cleaning stage. 
     For this we studied the $\chi^2$ statistic of the mean amplitude and spectral parameter
     maps output by the component separation code. The results are shown in Table
     \ref{tab:chi2}. For sky areas of 4000 sq deg or less, we find an acceptable probability
     for the incorrect model, with PTE of 7-10\%, and a bias of about $1\sigma$ in the
     estimated value for $r$. Larger sky areas give a larger bias, but here the problem would
     be more likely identified as the PTE is only $\sim1-2\%$. At this stage in the
     foreground cleaning, steps would be taken to account for an unidentified sky component.  
     As noted by \cite{2016MNRAS.458.2032R}, however, there is no guarantee that a foreground
     bias in $r$ would be recognised by a map-space component separation algorithm, although
     other strategies, such as studying the isotropy of the recovered primordial $B$-modes
     would also help in identifying foreground residuals. Note also that, even though we find
     that the foreground bias for the smaller sky areas is below or around the $1\sigma$ error,
     statistical noise fluctuations could push it to higher values and induce a fake,
     low-significance detection of primordial $B$-modes. A more detailed study of polarized
     AME emission is therefore necessary in order to assess its potential impact on $B$-mode
     searches.
     
   \subsection{Results: correlated noise} \label{sssec:results_sims_corrnoise}
     \begin{figure}
       \centering
       \includegraphics[width=0.49\textwidth]{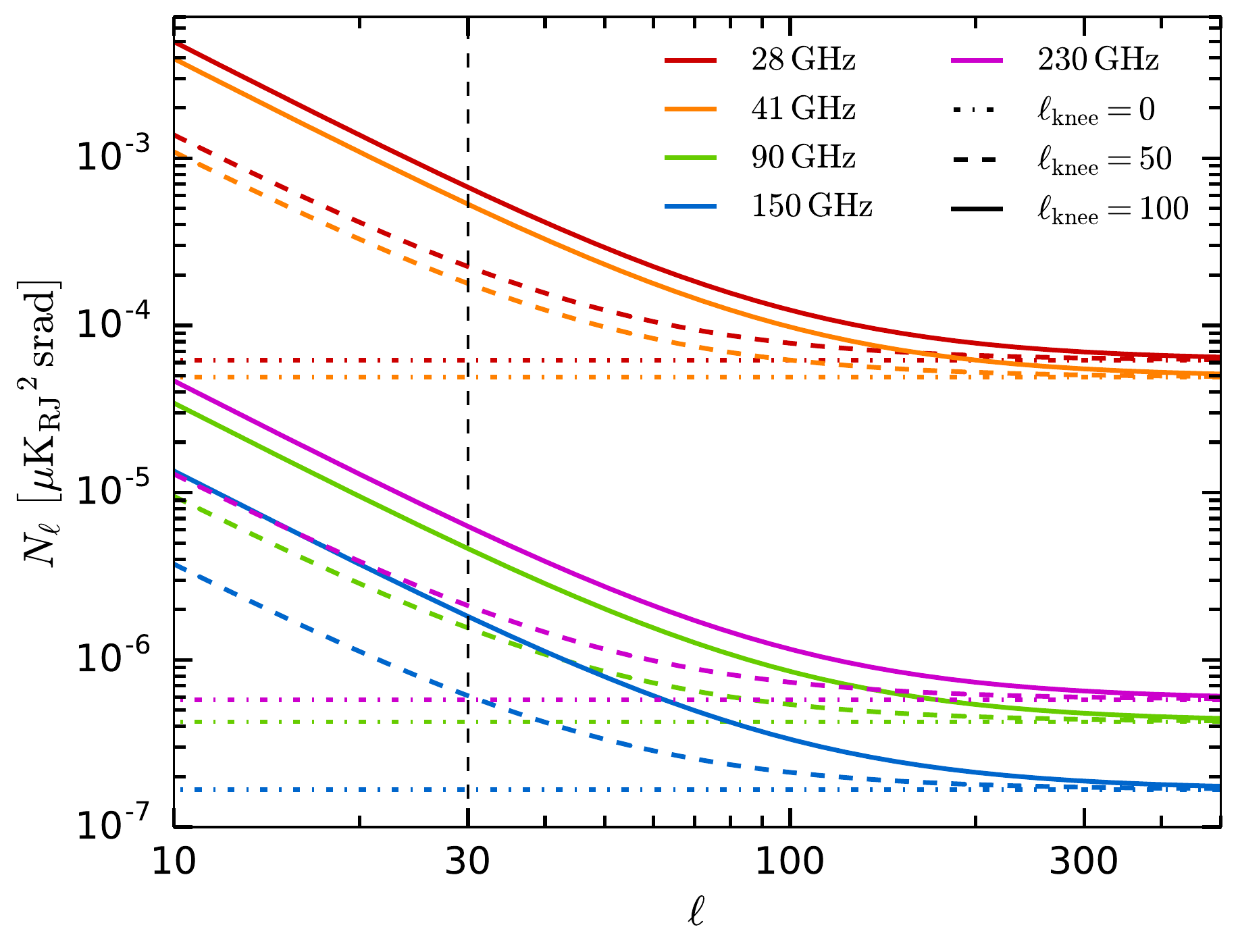}
       \caption{Noise curves assumed for S3 in the cases $\ell_{\rm knee}=100,\,50$ and 0
                (solid, dashed and dot-dashed lines). The dashed vertical line marks the smallest
                multipole included in the analysis in all cases.}
       \label{fig:sr_noises}
     \end{figure}
     \begin{figure}
       \centering
       \includegraphics[width=0.49\textwidth]{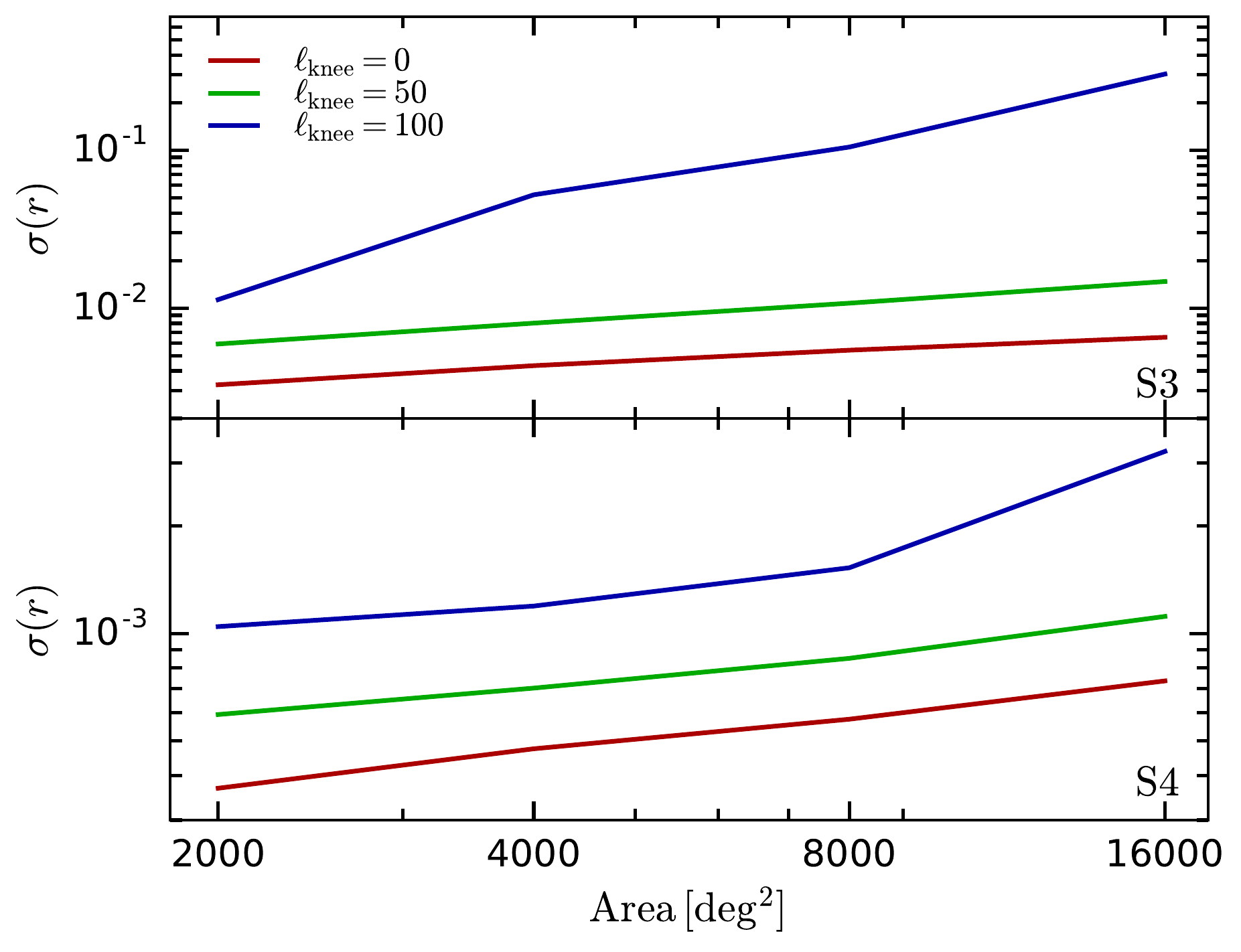}
       \caption{$68\%$ uncertainties on $r$ for S3 (upper panel) and S4 (lower panel)
                as a function of sky area and $\ell_{\rm knee}$, the scale above which the
                correlated noise component dominates. Results are shown for $\ell_{\rm knee}=0$
                (uncorrelated noise, red lines), $\ell_{\rm knee}=50$ and $\ell_{\rm knee}=100$
                (green and blue lines respectively). The sensitivity to $r$ degrades rapidly
                as the noise power on large scales increases.}
       \label{fig:sr_lknee}
     \end{figure}
     So far we have assumed ideal experiments characterized by a Gaussian beam and purely
     white noise. However, one of the main disadvantages of ground-based CMB experiments is the
     effect of atmospheric emission and other contamination, which can be correlated (non-white)
     on large scales.
     Although the properties of this atmospheric noise depend on the geographical location
     of the experiment, it will generically affect any measurement of large-scale observables,
     such as the signature of primordial tensor perturbations. In polarization, the effects of
     atmospheric noise can be mitigated instrumentally through the use of half-wave plates (HWP),
     which efficiently separate the polarized sky signal from the unpolarized atmospheric noise.
     Therefore it is important to explore the impact of correlated atmospheric noise on the
     final uncertainties in order to translate the science requirements into instrument
     specifications.
     
     In order to do this we repeated the analysis of our fiducial simulations (single thermal
     dust component and $0\%$ polarized AME) now including correlated noise. For this we used the
     model described in Section \ref{ssec:method_sims}, characterized by the parameter
     $\ell_{\rm knee}$, which determines the scale above which the noise becomes dominated by the
     correlated component. As shown in the left panel of Figure \ref{fig:fisher}, the optimal
     range of scales for constraining B-modes accessible to ground-based experiments is
     $\ell\lesssim100$, and therefore we studied the cases $\ell_{\rm knee}=50$ and
     $\ell_{\rm knee}=100$ (we will also show results for $\ell_{\rm knee}=0$, corresponding
     to the white-noise case studied before). Figure \ref{fig:sr_noises} shows examples of
     the noise power spectra used for S3 in this analysis. Note that, as before, in all
     cases we only used multipoles $\ell>30$ in the analysis.
     
     The results are shown in Figure \ref{fig:sr_lknee}. Not surprisingly, the final constraints
     on $r$ are sensitive to the level of large-scale noise, with the uncertainties
     increasing by factors of $\sim5$ and $\sim3$ for S3 and S4 respectively between the
     $\ell_{\rm knee}=0$ and $\ell_{\rm knee}=100$ cases. Under the requirement that S3 and 
     S4 experiments should be able to constrain the tensor-to-scalar ratio to better than
     $\sigma(r=0)\simeq10^{-2}$ and $10^{-3}$ respectively for a $10\%$ sky fraction, we
     estimate that the lowest multipole below which large-scale atmospheric noise can
     be allowed to dominate should be approximately $\ell^{\rm max}_{\rm knee}\sim50$.     

     \subsection{Results: comparison with other methods} \label{sssec:results_sims_ilc}
     \begin{figure}
       \centering
       \includegraphics[width=0.49\textwidth]{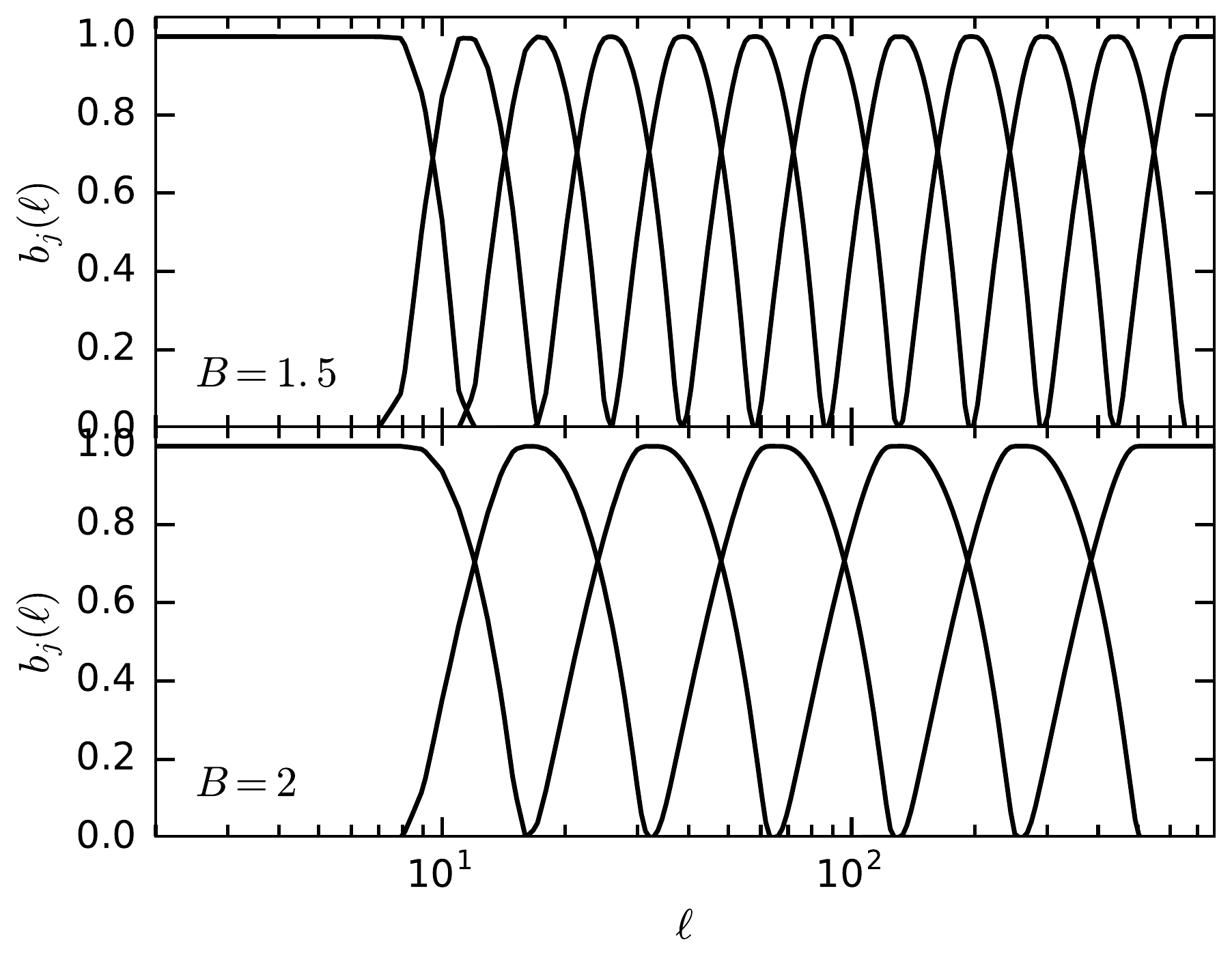}
       \caption{Window functions in harmonic space used to define the needlet basis used in the
                NILC algorithm. The upper and lower panels show the bands used for the $B$-adic
                parameter $B=1.5$ and $B=2$ respectively.}
       \label{fig:bands_nilc}
     \end{figure}
     \begin{figure}
       \centering
       \includegraphics[width=0.49\textwidth]{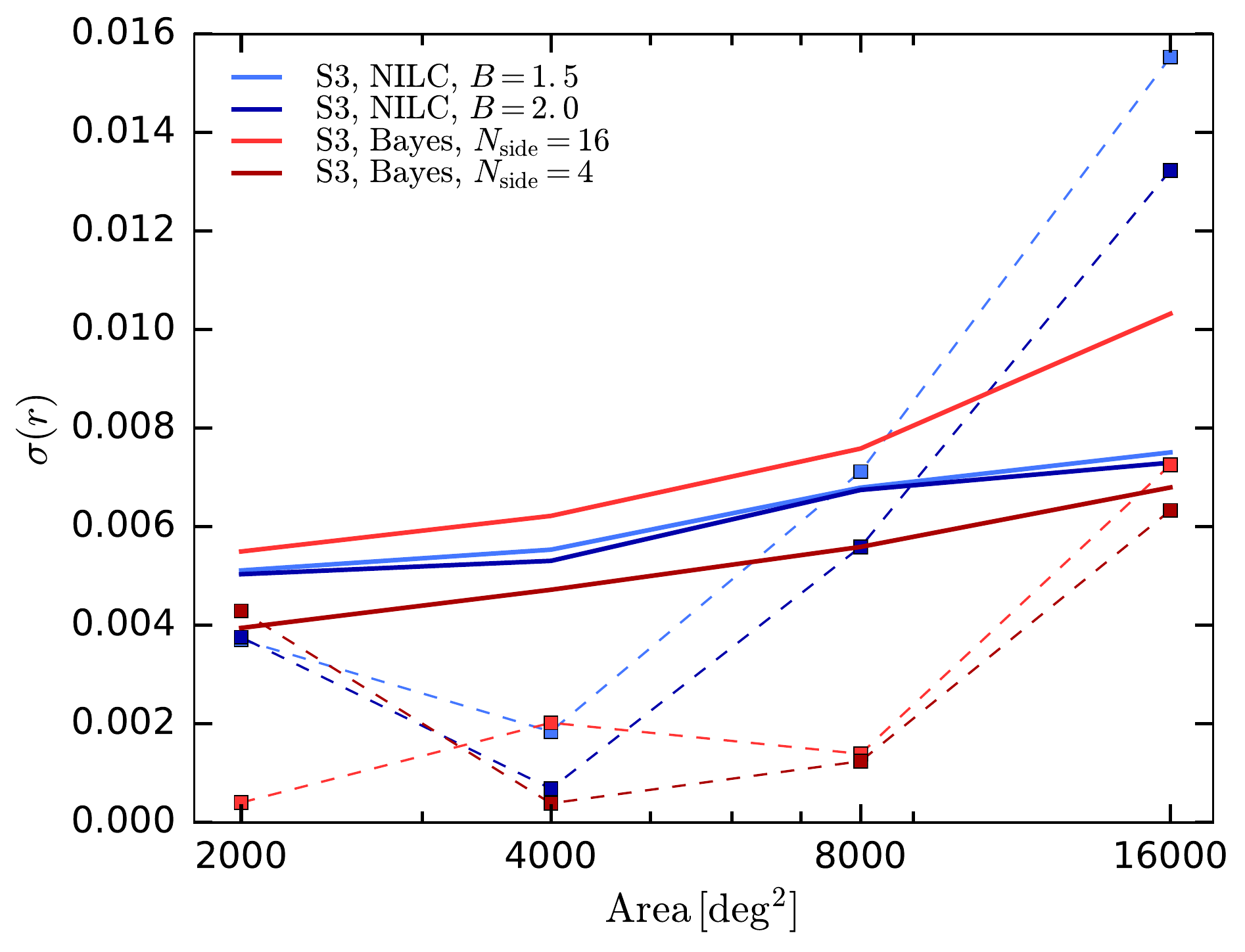}
       \caption{Results for S3 obtained using the Bayesian component separation code described
                in Section \ref{ssec:method_bayes} for spectral indices assumed constant in
                pixels of size $N_{\rm side}=4$ and 16 (light and dark red lines) compared with
                the results for a NILC algorithm defined using the $B$-adic mechanism for
                $B=1.5$ and 2 (light and dark blue lines). Solid lines show the 68\%-level
                uncertainties, while dashed lines show the best-fit values of $r$ in all
                cases. Similar uncertainties are found with both methods, with the Bayesian
                approach being the most optimal one when only a small number of spectral 
                indices are assumed free. A slight bias is found for the NILC algorithm 
                for the largest sky area.}
       \label{fig:sr_nilc}
     \end{figure}
     Many different methods have been used in the past to tackle the problem of component
     separation, and it would be interesting to study whether the results obtained here
     concerning the detectability of primordial $B$-modes are qualitatively universal
     across methods. To this end we have also implemented an independent version of the
     Needlet Internal Linear Combination algorithm (NILC, \cite{2009A&A...493..835D})
     and compared its results with those of the Bayesian approach described above. A key
     difference between both methods is the complete model-independence of NILC, which
     does not assume a specific spectral dependence for any components other than the one
     we wish to separate (in this case the CMB). This should make NILC more robust to
     badly modelled foreground components, at the cost of sub-optimal final uncertainties.
     On the other hand, a key drawback of ILC methods is that the noise level of the
     output maps, as well as a potential bias in them, must be measured using simulations.
     However, for the purposes of verifying the validity of our results in terms of the
     expected uncertainties on the tensor-to-scala ratio, NILC is an appropriate
     alternative algorithm to compare with.
     
     The details of the NILC algorithm have been thoroughly described in the literature
     \cite{2007arXiv0706.2598G,2010ApJ...723....1P}, and we will only describe the method
     briefly here. It consists of three main steps:
     \begin{itemize}
       \item All the maps in the different frequency bands are first decomposed into a
             set of needlet coefficients $\psi_j(\nv,\nu)$. These can be thought of as
             band-limited versions of the original maps in a set of multipole bands
             $b_j(\ell)$ characterized by a scale index $j$.             
       \item For each scale, an internal linear combination of the different frequency
             channels that extracts the CMB component is determined for each pixel using
             information from all other pixels in a disc around it. The size of this
             disc is chosen such that the number of independent modes in the disc is
             large enough to ensure a reliable determination of the frequency-frequency
             covariance matrix.
       \item After applying the internal linear combination, we are left with a set of
             foreground-cleaned needlet coefficient $\psi^c_j(\nv)$, which are then
             synthesized to generate the final cleaned CMB map.
     \end{itemize}
     In practice, we generalize this method to make use of polarized data by first
     transforming the input $(T,Q,U)$ maps at each frequency into the (pseudo)scalars
     $(T,E,B)$ and then applying the algorithm above to each component separately.
     
     As described above, the needlet transforms needed to carry out the NILC algorithm are
     defined by the set of multipole bands $b_j(\ell)$. In our implementation we define these
     functions through the so-called $B$-adic mechanism \cite{2007arXiv0706.2598G}. In this
     case all the bands are defined in terms of a single function $h(x)$ with support in the
     range $x\in[1/B,B]$. The band functions are then defined as $b_j(\ell)=h(\ell/B^j)$ and
     therefore have support in $\ell\in[B^{j-1},B^{j+1}]$. Thus, the spectral resolution of
     these bands is determined by the choice of $B>1$, and can be increased by choosing
     values of $B$ closer to $1$. The specific choice of $h(x)$ used in our implementation
     uses the guidelines of \cite{2010ApJ...723....1P}, and we studied the results for
     $B=1.5$ and 2. The corresponding $\ell$-bands for both cases are shown in Fig.
     \ref{fig:bands_nilc}.
     
     The result of this exercise is shown in Figure \ref{fig:sr_nilc} for S3 as a function of
     sky area. We see that, overall, we obtain uncertainties on $r$ that are somewhat larger
     than those obtained by the Bayesian component separation algorithm in the optimal
     $N_{\rm side}=4$ case. This reinforces our confidence in the expected uncertainties
     reported above. Furthermore, we observe a slight bias in the measured value of $r$
     on the largest sky areas (still below $2\sigma$), caused by the inability of the NILC
     method to fully remove the large-scale foregrounds when more contaminated regions
     are included in the analysis.

\section{Discussion}\label{sec:discuss}
  The detection of degree-scale CMB $B$-modes is one of the most important science cases for
  current and future CMB experiments, given the wealth of information contained in this
  observable. The detection and precise measurement of the $B$-mode power spectrum, however,
  is only achievable after accurately separating the cosmological signal from the large Galactic
  foregrounds in which it is immersed. In this work we have studied the ability of current
  (Stage-3) and future (Stage-4) ground-based experiments to recover the primordial
  $B$-mode signal in the presence of uncertain foregrounds. For this, we made use of
  realistic sky simulations processed through a data-analysis pipeline consisting of a
  Bayesian map-space component-separation code\footnote{The component-separation code
  is publicly available at \url{https://github.com/damonge/BFoRe}} and a power-spectrum
  estimator\footnote{The pseudo-$C_\ell$ estimator used here is publicly available at
  \url{https://github.com/damonge/NaMaster}}.
  We thus try to subject the data to the analysis methods that would be used in a realistic
  setting, which allows us to fully propagate foreground uncertainties into the final
  constraints on the tensor-to-scalar ratio, as well as to identify possible biases in those
  constraints caused by an incorrect foreground modelling.
  
  We find that accounting for highly resolved spectral indices causes a higher noise variance
  in the foreground-clean maps, which translates into larger final uncertainties on $r$. We
  therefore optimize the size of the resolution elements over which foreground spectral parameters
  are allowed to vary as a compromise between the size of the final error bars and the foreground
  bias associated with a poor representation of the spatial variation of these parameters. After
  doing that, we find that a Stage-3 AdvACT-like experiment should be able to constrain the
  tensor-to-scalar ratio to the level of $\sigma(r=0|{\rm S3})\simeq(4-6)\times10^{-3}$, with S4
  achieving sensitivities an order of magnitude better
  ($\sigma(r=0|{\rm S4})\simeq(4-6)\times10^{-4}$). These estimates are within the context of
  the foreground models considered in this analysis, which are consistent with current data but do
  not necessarily capture all possible scenarios. They also assume full mode recovery for a small
  sky patch, and Gaussian noise.

  Given the noise levels expected for S3, we find that it is always advantageous to push for deeper
  rather than wider observations if $r=0$, while the same is true for S4 only assuming optimal
  delensing levels. This is consistent with the findings of similar studies. We also find that
  delensing does not lead to a significant reduction in the forecast error on $r$ for S3, since
  the amplitude of the cosmic variance contribution from the lensing B-modes is comparable to the
  irreducible noise variance. Likewise, we observe that, given the noise levels of the 353 GHz Planck
  channel, its usefulness would be marginal for S3, and completely negligible for S4.
  
  We have also studied the effect of atmospheric noise, modelling it as a correlated large-scale
  component that dominates on scales $\ell$ below some $\ell_{\rm knee}$. We find that large-scale
  non-white noise can dramatically affect the final constraints on $r$, especially if it dominates
  on scales $\ell\lesssim100$, where most of the primordial $B$-mode signal is concentrated.
  We find that the uncertainties for purely white noise ($\ell_{\rm knee}=0$) grow by a factor of
  $\sim3$ and $\sim5$ for S4 and S3 respectively after assuming a large-scale correlated noise
  component dominating below $\ell_{\rm knee}=100$. Assuming that large-scale atmospheric noise can
  be kept under control, for instance through the use of half-wave plates, Stage-3 and Stage-4
  experiments should still be able to measure the tensor-to-scalar ratio with accuracies 
  $\sigma(r=0)=10^{-2}$ and $10^{-3}$ respectively, given the foreground models that we have
  considered. These measurements would directly translate into interesting constraints on
  inflationary theories.
  
  Given the currently large uncertainties in the nature of polarized diffuse foregrounds, it is
  important to study departures from the fiducial synchrotron $+$ single thermal dust model, in
  order to explore the possible biases on $r$ caused by incorrect foreground modelling. To this
  extent we have considered simulations containing two different polarized thermal dust components
  and a polarized AME component at low frequencies (modelled as polarized spinning dust emission).
  We find that, given the noise levels of both S3 and S4, a single-component dust model should be
  able to describe the 2-component model sufficiently well, with no detectable foreground bias on
  $r$. On the other hand, a $2\%$ polarized AME component would induce, if
  unaccounted for, a significant bias on $r$ for S4, although its effects would be negligible given
  the larger noise levels of S3. This highlights the importance of future measurements of the
  polarized sky at low frequencies (by e.g. \cite{2010ASSP...14..127R,2010SPIE.7741E..1IK}) in
  order to reduce our current uncertainties on the impact of polarized AME. There are also a
  variety of ways in which the true sky could be more complicated than any of the models we have
  considered here, and this will be the subject of future studies.
  
  It is also important to note that there are a number of potential sources of instrumental
  systematic uncertainties that we have not considered in this paper, such as
  temperature-to-polarization leakage, beam asymmetries or ground and scan-synchronous pick-up,
  which could impact the final constraints on $r$. In future work we also anticipate
  comparing forecasts for a given S4-type experimental configuration using our methods, with the
  methods described in \cite{2016JCAP...03..052E,Buza.inprep} as the definition of S4 is refined.
  
  In order to obtain reliable constraints on the amplitude of large-scale $B$-mode fluctuations
  with future sensitive ground-based facilities, large efforts are needed both in understanding
  the physics of diffuse polarized Galactic foreground and in designing experiments and data
  analysis methods able to separate these foregrounds from the cosmological CMB signal. To this
  extent, map-based component separation methods are able to consistently propagate foreground
  uncertainties caused by spatially varying spectral parameters, and, as we have shown, can
  provide clear diagnostics of incomplete foreground cleaning. These, together with suites of
  null tests aimed at identifying sources of astrophysical or instrumental systematic effects,
  provide a path towards placing robust constraints on the physics of the inflationary Universe.

\section*{Acknowledgments}
  We thank Victor Buza, John Carlstrom, Steve Choi, Josquin Errard, Stephen Feeney,
  John Kovac, Thibaut Louis, Lyman Page, Blake Sherwin and David Spergel for useful
  comments and discussions. DA recieves support from BIPAC, DA and JD are supported
  by ERC grant 259505 and BT acknowledges the support of an STFC studentship. We
  acknowledge the use of the Healpix software package.
 
\bibliography{paper}

\appendix
\begin{widetext}
  \section{Sampling the posterior}\label{app:methods}
    Here we describe the strategies used to sample the posterior distribution introduced
    in Section \ref{ssec:method_bayes}.
    \subsection{Sampling the posterior I: Gibbs sampling} \label{sssec:method_bayes_gibbs}
      The Gibbs sampling algorithm tries to solve the problem of sampling from a
      multivariate distribution by alternatively sampling from the distribution of
      the different parameters (or sets of parameters) conditional on the previous
      iterations of the rest. In our case, let ${\bf T}_i$, ${\bf b}_i$ denote the
      $i$-th sample of the amplitudes and spectral indices. We then draw the
      $(i+1)$-th sample as
      \begin{align}\label{eq:gibbs_amplitude}
        &{\bf T}_{i+1}\leftarrow p_l({\bf T}|{\bf d},{\bf b}_i)
                                 \propto p({\bf d}|{\bf T},{\bf b}_i)\,p_p({\bf T}),\\
        &{\bf b}_{i+1}\leftarrow p_l({\bf b}|{\bf d},{\bf T}_{i+1})
                                 \propto p({\bf d}|{\bf T}_{i+1},{\bf b})\,p_p({\bf b})
      \end{align}
      The advantage of this method is that, since the amplitudes are Gaussianly
      distributed, they can be sampled analytically with a 100\% acceptance ratio,
      thus gaining an enormous speed-up factor with respect to a naive Monte
      Carlo sampling of the individual parameters.
      
      Explicitly, the conditional distribution for ${\bf T}$ assuming no prior on the
      amplitudes can be written as
      \begin{equation}\label{eq:gauss_conditional}
        p({\bf T}|{\bf d},{\bf b})\propto\exp\left[-\frac{1}{2}
        \left({\bf T}-\bar{\bf T}\right)^T\hat{N}_T^{-1}
        \left({\bf T}-\bar{\bf T}\right)\right],
      \end{equation}
      where
      \begin{align}
        \hat{N}_T^{-1}=\hat{F}^T\hat{N}^{-1}\hat{F},\hspace{12pt}
        \bar{\bf T}=\hat{N}_T\,\left(\hat{F}^T\hat{N}^{-1}{\bf d}\right).
      \end{align}
      Samples from Eq. \ref{eq:gauss_conditional} can then be easily drawn as
      \begin{equation}
        {\bf T}_i=\bar{\bf T}+\hat{L}^{-1}\,{\bf u},
      \end{equation}
      where ${\bf u}$ is an uncorrelated, unit-variance Gaussian random vector, and
      $\hat{L}$ is the Choleski decomposition of $\hat{N}_T^{-1}$ (i.e.
      $\hat{N}_T^{-1}=\hat{L}^T\hat{L}$).
      The spectral indices are then sampled jointly using a multi-dimensional
      Monte-Carlo Markov Chain (MCMC) Metropolis-Hastings algorithm.

    \subsection{Sampling the posterior II: marginalising over amplitudes}
      \begin{figure*}
        \centering
        \includegraphics[width=0.49\textwidth]{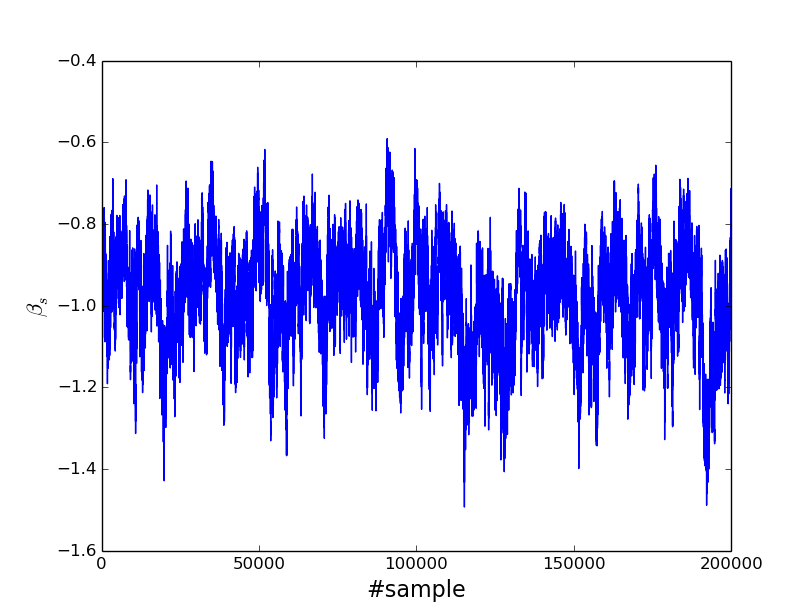}
        \includegraphics[width=0.49\textwidth]{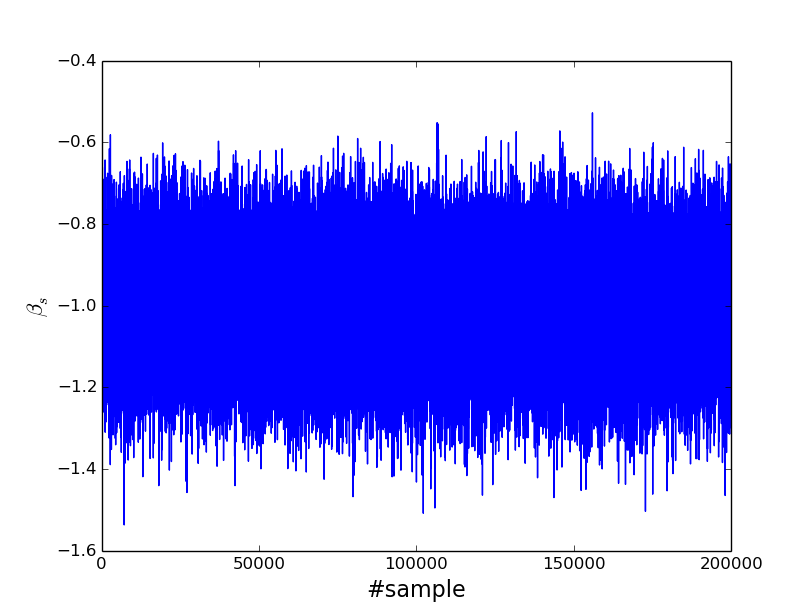}
        \caption{{\sl Left panel:} MCMC chain for $\beta_s$ in the case where
                 amplitudes are sampled via Gibbs sampling.
                 {\sl Right panel:} MCMC chain for $\beta_s$ for indices sampled
                 directly from the marginal distribution.}
        \label{fig:chains}
      \end{figure*}
      \label{sssec:method_bayes_marg}
      Since the amplitudes ${\bf T}$ are Gaussianly distributed, it is also possible to
      analytically marginalize over them to obtain the marginal distribution for the
      spectral indices. As noted above, the likelihood $p_l({\bf d}|{\bf T},{\bf b})$
      in Eq. \ref{eq:like1} can be written as in Eq. \ref{eq:gauss_conditional}.
      Writing the ${\bf b}$-dependent proportionality constant explicitly we obtain:
      \begin{align}\nonumber
        p({\bf T},{\bf b}|{\bf d})\propto&
        \frac{\exp\left[-\frac{1}{2}\left({\bf T}-\bar{\bf T}\right)^T
        \hat{N}_T^{-1}\left({\bf T}-\bar{\bf T}\right)^T\right]}
        {\sqrt{{\rm det}(\hat{N}_T)}}
        \times\\
        &\sqrt{{\rm det}(\hat{N}_T)}
        \exp\left[\frac{1}{2}\bar{\bf T}^T\hat{N}_T^{-1}\bar{\bf T}\right]p_p({\bf b}).
      \end{align}
      Integrating this equation over the amplitudes we see that the first term
      integrates out to a constant factor, and thus we obtain the marginal distribution
      for the spectral indices:
      \begin{equation}\label{eq:margdist}
        p({\bf b}|{\bf d})\propto\sqrt{{\rm det}(\hat{N}_T)}
        \exp\left[\frac{1}{2}\bar{\bf T}^T\hat{N}_T^{-1}\bar{\bf T}\right]p_p({\bf b}).
      \end{equation}

      Why is it relevant to go through all this trouble? Ultimately we are interested
      in the moments of the distribution of the amplitude of the CMB component. The
      expectation value for any function of the amplitudes can be computed as an
      integral over the marginal distribution:
      \begin{align}
        \langle g({\bf T})|{\bf d}\rangle
        &=\int d{\bf T}d{\bf b}\, g({\bf T})p({\bf T},{\bf b}|{\bf d}),\\
        &=\int d{\bf b}\,\langle g({\bf T})|{\bf b},{\bf d}\rangle\, p({\bf b}|{\bf d})
      \end{align}
      Since the conditional distribution $p({\bf T}|{\bf d},{\bf b})$ is Gaussian, it is
      completely defined by the first 2 moments of the distribution:
      \begin{align}
        \langle{\bf T}|{\bf d},{\bf b}\rangle=\bar{\bf T},\hspace{12pt}
        \langle({\bf T}-\bar{\bf T})({\bf T}-\bar{\bf T})^T|{\bf d},{\bf b}\rangle=\hat{N}_T,
      \end{align}
      which can be computed analytically for any value of ${\bf b}$. Thus we can compute the
      marginalized mean and covariance of the amplitudes by sampling only the spectral
      indices from their marginal posterior in Eq. \ref{eq:margdist} and computing the
      analytical conditional mean and covariance for each sample.
      
      It is easy to see how doing this would improve the performance of the method. By
      skipping the intermediate sampling of the amplitudes (Eq. \ref{eq:gibbs_amplitude}),
      we reduce the correlation length of the the MCMC chains for ${\bf b}$, and fewer
      samples are needed to cover the posterior. Figure \ref{fig:chains} shows a
      Monte-Carlo chain for $\beta_s$ in the full Gibbs-sampling scheme of the previous
      section (left) and sampling directly from the marginal distribution (right).

\section{Volume prior for spectral parameters}\label{app:volume}
  \begin{figure}
    \centering
    \includegraphics[width=0.49\textwidth]{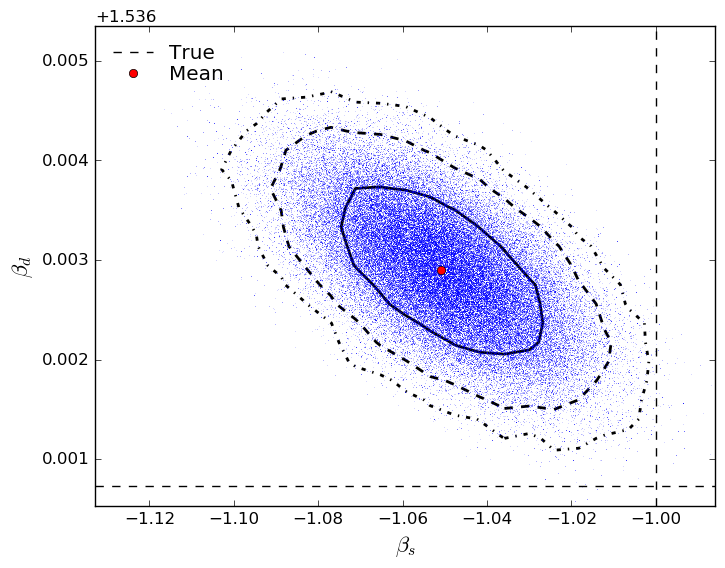}
    \includegraphics[width=0.49\textwidth]{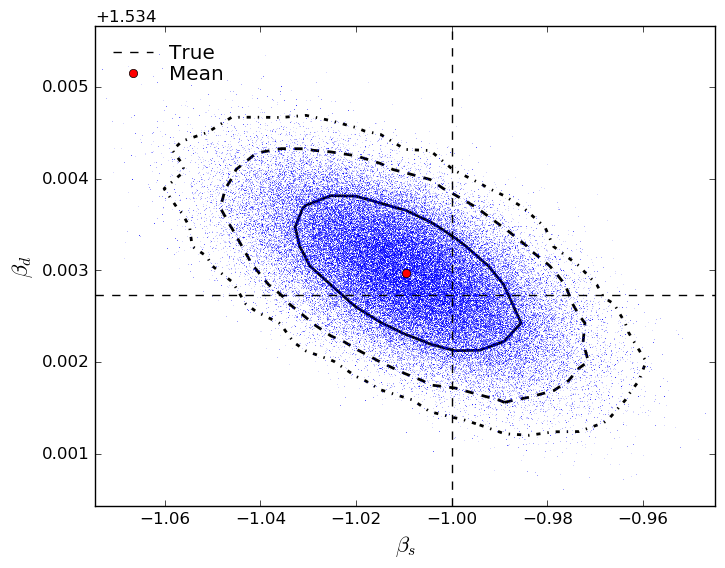}
    \includegraphics[width=0.49\textwidth]{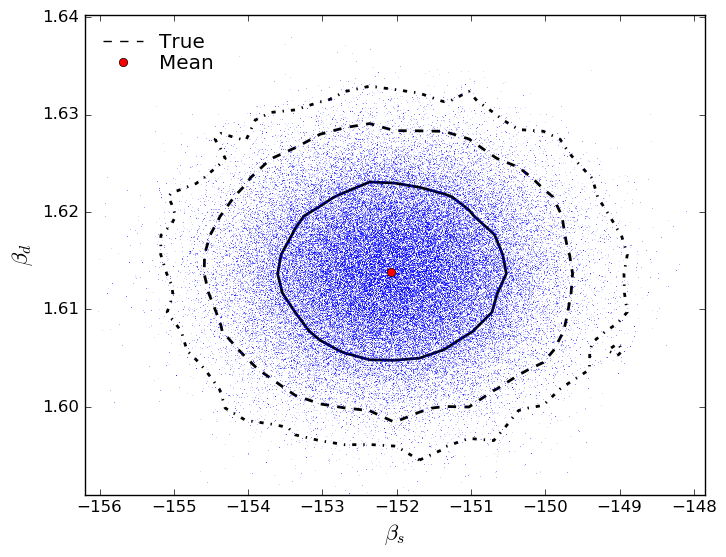}
    \includegraphics[width=0.49\textwidth]{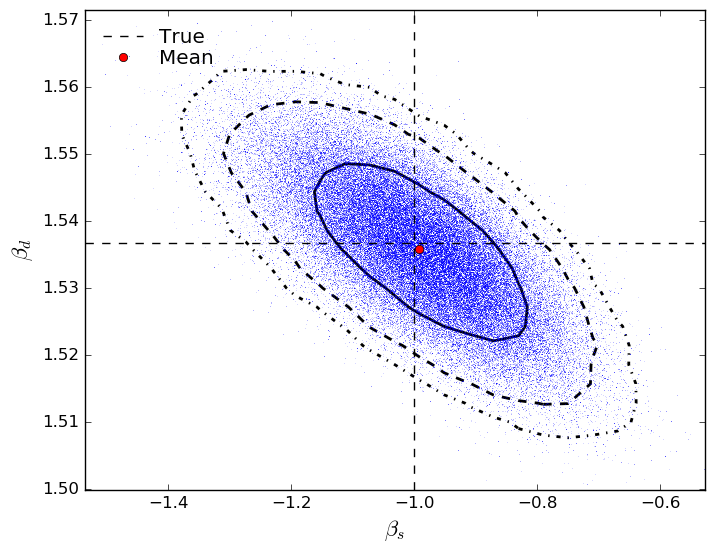}
    \caption{{\sl Upper left panel:} contour plots ($1,2,3\sigma$ - solid, dashed and
             dot-dashed lines), mean value (red circle) and true value (dashed lines)
             of the dust and synchrotron spectral indices for a high-SNR region without
             cancelling the volume factor. The posterior mean is more than 3-sigmas
             away from the true input value. {\sl Upper right panel:} same plot after
             cancelling the volume prior. The true value lies within $2\sigma$ of the
             posterior mean. The {\sl lower panels} show the same for a low-SNR region.}
             \label{fig:bias_volume_hi}
  \end{figure}
  As has been noted in the literature, when dealing with non-linear parameters, a flat
  prior is not necessarily appropriate to describe quantitatively our ignorance about
  their value. This can be easily ilustrated using the results of Appendix
  \ref{sssec:method_bayes_marg}.
  
  Consider first the case of a noise-dominated map, where we can approximate the data
  as being completely made up of noise. In that case, the expectation value of the
  exponent in Eq. \ref{eq:margdist} is given by:
  \begin{equation}
    \left\langle\bar{\bf T}\hat{N}_T^{-1}\bar{\bf T}\right\rangle=
    {\rm Tr}\left[\hat{N}_T\hat{F}^T\hat{N}^{-1}\langle{\bf d}{\bf d}^T\rangle
    \hat{N}^{-1}\hat{F}\right]=N_A,
  \end{equation}
  where $N_A$ is the number of amplitudes. Thus, in this regime, the posterior for
  the spectral parameters ${\bf b}$ is dominated by the prefactor $\sqrt{|\hat{N}_T|}$.
  In the simplified case where the only component is synchrotron, and in the absence
  of polarized channels, this volume factor is given by:
  \begin{equation}\label{eq:vol_synch}
    \sqrt{|\hat{N}_T|}\propto\left[\sum_\nu
    \left(\frac{\nu}{\nu_0^s}\right)^{2\beta_s}\frac{1}{\sigma_\nu^2}\right]^{-1/2},
  \end{equation}
  which becomes arbitrarily large for $\beta_s\rightarrow-\infty$. This result makes
  qualitative sense: in the absence of data, we must cancel the synchrotron component.
  This can be achieved by either having very small amplitudes or a very steep spectral
  index. Since a large $|\beta_s|$ would give rise to a negligible synchrotron
  component, even for very large amplitudes, this option covers a larger volume of the
  space of parameters.
  
  However, is this behavior desirable? Presumably, we would expect that, in the absence
  of signal, the spectral indices would be completely unconstrained, or else dominated
  by the prior $p_p({\bf b})$. More importantly, since this volume factor does not
  depend on the data, could its presence bias our estimate of ${\bf b}$ in the case
  where the data contains a measurable amount of signal? This can be easily shown to
  be the case \cite{2008ApJ...676...10E}. The upper panels of Figure
  \ref{fig:bias_volume_hi} show the likelihood contours for the dust and synchrotron
  spectral indices in a simulated patch of the sky with relatively high signal to
  noise. The left and right panels present the result before and after cancelling
  the volume factor respectively. The dashed lines show the true values of the
  spectral indices, and the red circles correspond to the mean of the posterior. The
  presence of the volume factor biases the estimate of the spectral indices by more
  than $4\sigma$, and their true value is recovered within $1\sigma$ after accounting
  for it. The situation in a low-$S/N$ region is ilustrated in the lower panels of
  Fig. \ref{fig:bias_volume_hi}: even using a broad Gaussian prior with
  $\sigma_\beta\sim1.$ centered on the true values of the spectral indices, the mean
  estimated indices are hundreds of $\sigma$s away from their true values. These,
  however, are well recovered after cancelling the volume factor and, as shown in
  the right panel of this Figure, their uncertainty is not dominated at all by the
  Gaussian prior.
      
  These two undesirable features (i.e. the non-flat posterior for the spectral indices
  in the signal-less case, and the bias it entails) can be avoided by including a
  factor $|\hat{N}_T|^{-1/2}$ in our prior that exactly cancels the volume factor.
  This problem can also be solved by applying a Jeffreys prior, given by the square
  root of the Fisher information matrix \cite{2008ApJ...676...10E}
  \begin{equation}
    p_{\rm Jef}(\theta)\propto\sqrt{F_{\theta\theta}}=
    \sqrt{-\left\langle\frac{\partial^2\mathcal{L}}{\partial\theta^2}\right\rangle}.
  \end{equation}
  Note however, that for the case of a power-law spectral index, this is given by
  \begin{equation}
    p_{\rm Jef}(\beta_s)\propto\sqrt{\sum_\nu
    \left[\left(\frac{\nu}{\nu_0^s}\right)^{\beta_s}
    \frac{1}{\sigma_v}\ln\left(\frac{\nu}{\nu_0^s}\right)\right]^2},
  \end{equation}
  which is equivalent to the volume prior $p_{\rm Vol}\equiv|\hat{N}_T|^{-1/2}$
  defined by Eq. \ref{eq:vol_synch} except for the subdominant logarithmic term.
\end{widetext}

\end{document}